\documentclass[12pt,preprint]{aastex}
\usepackage{epsf}
\def\nbody{$n$-body}
\def\deg{\ifmmode {^\circ}\else {$^\circ$}\fi}
\def\degree{\ifmmode {^\circ}\else {$^\circ$}\fi}
\def\mum{\ifmmode {\rm \,\mu {\rm m}}\else $\rm \,\mu {\rm m}$\fi}
\def\arcsec{\ifmmode ^{\prime \prime}\else $^{\prime \prime}$\fi}
\def\secpoint{\mbox{$''\mskip-7.6mu.\,$}}
\def\minpoint{\mbox{$'\mskip-4.6mu.$}}

\def\inch{\ifmmode ^{\prime \prime}\else $^{\prime \prime}$\fi}
\def\gs{\ifmmode {{\rm g~s^{-1}}}\else ${\rm g~s^{-1}}$\fi}
\def\msunyr{\ifmmode {M_{\odot}~{\rm yr^{-1}}}\else $M_{\odot}~{\rm yr^{-1}}$\fi}
\def\msun{\ifmmode {M_{\odot}}\else $M_{\odot}$\fi}
\def\rsun{\ifmmode {R_{\odot}}\else $R_{\odot}$\fi}
\def\lsun{\ifmmode {L_{\odot}}\else $L_{\odot}$\fi}
\def\mstar{\ifmmode {M_{\star}}\else $M_{\star}$\fi}
\def\rstar{\ifmmode {R_{\star}}\else $R_{\star}$\fi}
\def\tstar{\ifmmode {T_{\star}}\else $T_{\star}$\fi}
\def\lstar{\ifmmode {L_{\star}}\else $L_{\star}$\fi}
\def\mwd{\ifmmode {M_{wd}}\else $M_{wd}$\fi}
\def\rwd{\ifmmode {R_{wd}}\else $R_{wd}$\fi}
\def\twd{\ifmmode {T_{wd}}\else $T_{wd}$\fi}
\def\lwd{\ifmmode {L_{wd}}\else $L_{wd}$\fi}
\def\md{\ifmmode {M_d}\else $M_d$\fi}
\def\ld{\ifmmode {L_d}\else $L_d$\fi}
\def\ad{\ifmmode A_d\else $A_d$\fi}
\def\ldlwd{\ifmmode L_d / L_{wd}\else $L_d / L_{wd}$\fi}
\def\ldlstar{\ifmmode L_d / L_\star\else $L_d / L_{\star}$\fi}
\def\rearth{\ifmmode {\rm R_{\oplus}}\else $\rm R_{\oplus}$\fi}
\def\mearth{\ifmmode {\rm M_{\oplus}}\else $\rm M_{\oplus}$\fi}
\def\qdstar{\ifmmode Q_D^\star\else $Q_D^\star$\fi}
\def\vsqd{\ifmmode v^2 / Q_D^\star\else $v^2 / Q_D^\star$\fi}
\def\kms{\ifmmode {\rm km~s^{-1}}\else $\rm km~s^{-1}$\fi}
\def\ms{\ifmmode {\rm m~s^{-1}}\else $\rm m~s^{-1}$\fi}
\def\vrel{\ifmmode v_{rel}\else $v_{rel}$\fi}
\def\mdot{\ifmmode \dot{M}\else $\dot{M}$\fi}
\def\mdotz{\ifmmode \dot{M}_0\else $\dot{M}_0$\fi}
\def\mesc{\ifmmode m_{esc}\else $m_{esc}$\fi}
\def\rmin{\ifmmode r_{min}\else $r_{min}$\fi}
\def\rmax{\ifmmode r_{max}\else $r_{max}$\fi}
\def\xmax{\ifmmode x_{max}\else $x_{max}$\fi}
\def\mmin{\ifmmode m_{min}\else $m_{min}$\fi}
\def\mmax{\ifmmode m_{max}\else $m_{max}$\fi}
\def\rmind{\ifmmode r_{min,d}\else $r_{min,d}$\fi}
\def\rmaxd{\ifmmode r_{max,d}\else $r_{max,d}$\fi}
\def\mmaxd{\ifmmode m_{max,d}\else $m_{max,d}$\fi}
\def\vrad{\ifmmode v_{rad}\else $v_{rad}$\fi}
\def\qz{\ifmmode q_{0}\else $q_{0}$\fi}
\def\qi{\ifmmode q_{i}\else $q_{i}$\fi}
\def\ql{\ifmmode q_{l}\else $q_{l}$\fi}
\def\qs{\ifmmode q_{s}\else $q_{s}$\fi}
\def\vhill{\ifmmode v_H\else $r_H$\fi}
\def\rhill{\ifmmode r_H\else $r_H$\fi}
\def\Rhill{\ifmmode R_H\else $R_H$\fi}
\def\rbrk{\ifmmode r_{brk}\else $r_{brk}$\fi}
\def\rdamp{\ifmmode r_{damp}\else $r_{damp}$\fi}
\def\rin{\ifmmode r_{in}\else $r_{in}$\fi}
\def\rout{\ifmmode r_{out}\else $r_{out}$\fi}
\def\tin{\ifmmode t_{in}\else $t_{in}$\fi}
\def\tout{\ifmmode t_{out}\else $t_{out}$\fi}
\def\ain{\ifmmode a_{in}\else $a_{in}$\fi}
\def\aout{\ifmmode a_{out}\else $a_{out}$\fi}
\def\r0{\ifmmode r_{0}\else $r_{0}$\fi}
\def\R0{\ifmmode R_{0}\else $R_{0}$\fi}
\def\m0{\ifmmode m_{0}\else $m_{0}$\fi}
\def\M0{\ifmmode M_{0}\else $M_{0}$\fi}
\def\xm{\ifmmode x_{m}\else $x_{m}$\fi}
\def\sigz{\ifmmode \Sigma_0\else $\Sigma_0$\fi}
\def\ergg{\ifmmode {\rm erg~g^{-1}}\else ${\rm erg~g^{-1}}$\fi}
\def\gyr{\ifmmode {\rm g~yr^{-1}}\else ${\rm g~yr^{-1}}$\fi}
\def\cms{\ifmmode {\rm cm~s^{-1}}\else ${\rm cm~s^{-1}}$\fi}
\def\gcms{\ifmmode {\rm g~cm^{-2}}\else $\rm g~cm^{-2}$\fi}
\def\gcmc{\ifmmode {\rm g~cm^{-3}}\else $\rm g~cm^{-3}$\fi}
\def\atil{\ifmmode {\tilde{a}}\else $\tilde{a}$\fi}
\def\ttil{\ifmmode {\tilde{t}}\else $\tilde{t}$\fi}
\def\sqrttt{\ifmmode {\tilde{t}^{1/2}}\else $\tilde{t}^{1/2}$\fi}

\def\orch{{\it Orchestra}}
\def\nh{{\it New Horizons}}
\def\pc{Pluto--Charon}
\def\mp{\ifmmode m_P\else $m_P$\fi}
\def\mc{\ifmmode m_C\else $m_C$\fi}
\def\mh{\ifmmode m_H\else $m_H$\fi}
\def\mk{\ifmmode m_K\else $m_K$\fi}
\def\ms{\ifmmode m_S\else $m_S$\fi}
\def\mn{\ifmmode m_N\else $m_N$\fi}
\def\rp{\ifmmode r_P\else $r_P$\fi}
\def\rc{\ifmmode r_C\else $r_C$\fi}
\def\apc{\ifmmode a_{PC}\else $a_{PC}$\fi}
\def\mpc{\ifmmode m_{PC}\else $m_{PC}$\fi}
\def\epc{\ifmmode e_{PC}\else $e_{PC}$\fi}

\begin{document}

\title{A \pc\ Sonata: The Dynamical Architecture of the Circumbinary Satellite System}
\vskip 7ex
\author{Scott J. Kenyon}
\affil{Smithsonian Astrophysical Observatory,
60 Garden Street, Cambridge, MA 02138}
\email{e-mail: skenyon@cfa.harvard.edu}

\author{Benjamin C. Bromley}
\affil{Department of Physics \& Astronomy, University of Utah,
201 JFB, Salt Lake City, UT 84112}
\email{e-mail: bromley@physics.utah.edu}

\begin{abstract}

Using a large suite of \nbody\ simulations, we explore the discovery space
for new satellites in the \pc\ system. For the adopted masses and orbits of 
the known satellites, there are few stable prograde or polar orbits with 
semimajor axes $a \lesssim 1.1~a_H$, where $a_H$ is the semimajor axis of 
the outermost moon Hydra. Small moons with radii $r \lesssim$ 2~km and 
$a \lesssim 1.1~a_H$ are ejected on time scales ranging from several yr 
to more than 10~Myr.  Orbits with $a \gtrsim 1.1~a_H$ are stable on time 
scales exceeding 100~Myr.  Near-IR and mid-IR imaging with JWST and 
ground-based occultation campaigns with 2--3-m class telescopes can detect 
1--2~km satellites outside the orbit of Hydra. Searches for these moons 
enable new constraints on the masses of the known satellites and on theories 
for circumbinary satellite formation.

\end{abstract}

\keywords{
planets and satellites: dynamical evolution and stability ---
planets and satellites: individual (Pluto)
}

\section{INTRODUCTION}
\label{sec: intro}

With four small satellites orbiting a binary planet, the Pluto-Charon system 
is a dynamical wonder \citep{weaver2006,buie2006,tholen2008,showalter2011,
youdin2012,buie2012,showalter2012,buie2013,brozovic2015,showalter2015}.  
The smallest satellite, Styx, lies reasonably close to the innermost stable 
orbit of the central binary.  The larger satellites -- Nix, Kerberos, and 
Hydra -- are packed about as tightly as possible.  Although the orbital 
periods of the satellites are almost integer multiples of the \pc\ period, 
the satellites rotate chaotically on time scales much shorter than their 
orbital periods.

Spectacular images and spectroscopic data from the \nh\ flyby provide new 
insights \citep[e.g.,][]{bagenal2016,grundy2016,weaver2016,mckinnon2017,
robbins2017,verbiscer2018,lauer2018}.  
The satellites are irregularly shaped, with equivalent spherical radii ranging 
from 10--12~km for Styx and Kerberos to 19--21~km for Nix and Hydra \citep[see 
also][]{showalter2015}.  Confirming earlier predictions \citep{youdin2012}, 
albedos are large, ranging from roughly 55\% for Kerberos and Nix to 65\% for 
Styx to nearly 85\% for Hydra.  Thus, the surfaces of the satellites are much 
icier than Pluto or Charon \citep{cook2018}. Expanding on the results of 
\citet{showalter2015}, detailed analysis by the \nh\ team shows that the 
satellites are not tidally locked with the central binary; rotational periods 
range from 0.43~d for Hydra to 5.31~d for Kerberos. 

Confounding theoretical expectations \citep[e.g.,][]{kb2014a,bk2015b}, \nh\ did
not detect any new satellites.  Data from a deep imaging survey place an upper 
limit on the radius, $r \lesssim$ 1.7~km, for smaller icy moons within 80,000~km 
($a \lesssim 1.23~a_H$) of the \pc\ center-of-mass \citep{weaver2016}.  This 
result is a weak test of numerical simulations for satellite formation, which 
predicted several small satellites, $r \lesssim$ 1--3~km, within 105,000~km of 
\pc\ ($a \approx 1.2 - 1.6~a_H$). The \nh\ data also place strong constraints on 
dusty debris orbiting \pc\ in the vicinity of the known satellites 
\citep{bagenal2016,lauer2018}. The observed upper limit on the optical depth, 
$\tau \lesssim 10^{-8}$ on $10^3 - 10^4$~km scales, is well below early 
predictions of $\tau \approx 10^{-7} - 10^{-5}$ \citep[e.g.,][]{stern2006,
steffl2007,poppe2011} but above more recent predictions of 
$\tau \approx 10^{-11}$ \citep{pires2013}.

Without another \pc\ flyby in the near future, placing additional constraints 
on the properties of the satellite system requires (i) more detailed analyses 
of existing data, (ii) expanded sets of numerical simulations, and (iii) new
high quality observations from the ground or from a near-Earth telescope.  In 
their discovery paper for Styx, \citet{showalter2012} note that this moon is 
close to the detection limit for HST.  Thus, it seems unlikely HST will find 
any fainter satellites \citep[see also][]{steffl2006}.  

To isolate the possible discovery space for new satellites, we conduct an 
extensive series of \nbody\ simulations. Our calculations follow the orbits 
of massive proxies for the four known satellites and sets of massless tracer 
particles orbiting a central binary with the physical properties of \pc. 
Adopting the masses in Table~\ref{tab: sats} and defining $a_S$ ($a_H$) as
the semimajor axis of Styx (Hydra), our results demonstrate that nearly all 
test particles with initial semimajor axes 
$0.95~a_S \lesssim a_0 \lesssim 1.1~a_H$ are unstable on time scales 
$t \lesssim$ 10--20~Myr. Test particles with the most stable orbits have 
$a_0 \approx 0.85 - 0.93~a_S$ ($a_0 \approx$ 36,000--40,000~km) or 
$a_0 \gtrsim 1.1~a_H$ ($a_0 \gtrsim$ 75,000~km).  

Additional calculations show that massive satellites with radii $r \lesssim$ 
1--2~km and mass density $\rho \approx$ 1~\gcmc\ on orbits with $a \gtrsim$ 
75,000~km are stable on 100--150~Myr time scales. In this semimajor axis range,
polar and prograde orbits are comparably stable.  Nearly all small moons with
$0.8~a_S \lesssim a_0 \lesssim 1.1~a_H$ are ejected on time scales ranging from
a few yr to $\sim$ 100--150~Myr. Together with the calculations of massless 
tracers, these results clarify the most promising location for new satellites
in the \pc\ system: outside the orbit of Hydra.

We then consider two options for finding new satellites.  With HST precluded, we 
show that (i) near-IR imaging with modest integration times on JWST instruments 
and (ii) occultations on medium-sized ground-based telescopes can track the orbits, 
shapes, and reflective properties of the known satellites and discover possible 
smaller satellites with high albedo and radii of 1--2~km. 

After briefly describing the \nbody\ code (\S2.1) and the physical properties of 
the known satellites (\S2.2), we consider the stability of circumbinary satellites 
in systems without and with the known \pc\ satellites in \S2.3--\S2.5. We outline 
the observational programs in \S3.  We conclude with a brief discussion (\S4) and 
summary (\S5).

\section{NUMERICAL SIMULATIONS}
\label{sec: sims}

\subsection{Background}

To explore the extent of the plausible discovery space for new satellites in 
the \pc\ system, we consider numerical simulations with \orch, a parallel 
\verb!C++/MPI! hybrid coagulation + \nbody\ code that follows the accretion, 
fragmentation, and orbital evolution of solid particles ranging in size from 
a few microns to thousands of km \citep{kenyon2002,bk2006,kb2008,bk2011a,bk2013,
kb2016a,knb2016}.  The ensemble of codes within \orch\ includes a multi-annulus 
coagulation code, an \nbody\ code, and radial diffusion codes for solids and 
gas.  Several algorithms link the codes together, enabling each component to 
react to the evolution of other components.

Here, we use the \nbody\ code to track the orbits of massive satellites and many 
massless tracers around the \pc\ binary. For the satellites, we adopt the masses,
radii, and orbital elements listed in Table~\ref{tab: sats} \citep{brozovic2015,
showalter2015,weaver2016} and initial state vectors from Table 8 of 
\citet{brozovic2015}.  To derive the position and velocity of Pluto relative to 
the system barycenter, we adopt an orbital period of 6.387~d and 
\mp\ = $1.303 \times 10^{25}$~g, \rp\ = 1183~km, and $f_P \lesssim$ 0.006 
for the mass, radius, and oblateness of Pluto; Charon has mass \mc\ = 
$1.587 \times 10^{24}$~g = 0.12~\mp, radius \rc\ = 606~km = 0.51~\rp, and
oblateness $f_C \lesssim$ 0.005 \citep[e.g.,][and references therein]{young1994,
person2006,brozovic2015,stern2015,nimmo2017,mckinnon2017}. To simplify assigning
orbits for massless tracer particles, we rotate the cartesian coordinate system 
of \citet{brozovic2015} to place the angular momentum vector $L$ in the z-direction. 

In this study, we do not consider how errors in the measured positions and velocities
of \pc\ and the smaller satellites might impact outcomes of the calculations. For a
nominal distance of 40~AU from the Earth, a 0\secpoint01 uncertainty in the centroid
of the point spread function for a small satellite on an HST image corresponds to $\sim$ 
7~km.  Compared to the observed semimajor axes ($4.2 \times 10^4 - 6.5 \times 10^5$~km)
or the derived radii of their Hill spheres (200--700~km), this uncertainty is rather 
small. We performed several test calculations with initial positions differing by 
5--10~km from the nominal positions for the small satellites. The results are identical 
to those starting from the nominal initial state vector. Thus, we consider results only
for one initial state vector.

To evolve the ensemble of massless tracers, we place them on orbits with initial 
semimajor axis $a$, $e$, and $\imath$ relative to the binary center-of-mass. These 
orbits are similar to a set of `most circular' circumbinary orbits with identical
orbital elements \citep[e.g.,][]{lee2006,lith2008a,youdin2012,leung2013,bk2015a,
bk2015b,bk2017}.  Adding circumbinary satellites to the \pc\ binary limits the set
of stable orbits for massless tracers \citep{lith2008a,youdin2012}. Our goal is to 
find the set of stable tracer orbits for an adopted set of properties for the known 
satellites.

For each suite of simulations, we derive results using a symplectic integrator which 
divides the \pc\ orbit into $N$ steps and maintains a constant time step throughout 
the integration. As outlined in the Appendix, several tests demonstrate that a minimum 
$N$ = 40 enables calculations with negligible drift over 100--500~Myr in $a$ and $e$ 
for the \pc\ binary and for small satellites with their nominal masses.  The algorithm 
has also been verified with test simulations in previous papers 
\citep[e.g.,][]{dunc1998,bk2006,bk2011a,bk2011b,bk2013,bk2017}.

On the NASA `discover' computer system, we perform calculations on either 
1 processor (6-bodies or 7-bodies) or 56 processors (\pc\ binary with massless 
tracers, with or without the small satellites). With 1 processor, a system 
with \pc\ and the four known satellites evolves 4.2~Myr per cpu-day. Adding 
another small satellite reduces the evolution time to 3.0~Myr per cpu day. At 
these rates, completing a typical 100~Myr calculation requires 25--34 days. 
Multi-processor calculations with the central binary, $56 ~ \times ~ 56$ = 
3136 tracers, and the 4 small satellites complete 1.9~Myr of evolution per day. 
As the \pc\ system ejects tracers, the calculations move somewhat faster.  
Typical 10--15 Myr calculations for these systems finish in 4--8 days.

For computational convenience, we perform calculations of tracers over a small 
range in semimajor axis $a$. When the small satellites are included, the range
in $a$ covers regions from (i) well inside to just outside the orbit of Styx, 
(ii) just inside the orbit of one of the small satellites to just outside the
orbit of the next satellite with larger $a$, or 
(iii) just inside to well outside the orbit of Hydra.
Instead of having a uniform density of tracers with $a$, this procedure generates 
an overlap of tracers co-rotating with each small satellite. Aside from allowing
us to compare results for tracers with similar $a$ from different calculations, 
these starting conditions provide a better measure of the survival rate for 
co-rotating tracers.

\subsection{Physical Properties of the \pc\ Satellites}

To derive physical properties for the four small satellites (Table~\ref{tab: sats}), 
we rely on published observations and \nbody\ simulations. Orbital elements -- the
semimajor axis $a$, eccentricity $e$, inclination $i$, and orbital period $P_{orb}$ --
are from detailed analyses of HST imaging data \citep{brozovic2015,showalter2015}.

Comprehensive imaging data from HST and the \nh\ flyby demonstrate that all of the 
satellites have irregular, oblong shapes \citep{showalter2015,weaver2016}. Aspect 
ratios are roughly 2:1:1 (Styx and Kerberos), 1.5:1:1 (Nix), or 3:2:1 (Hydra). 
For numerical simulations, we adopt equivalent spherical radii 
$r_i = \sqrt[3]{x_i y_i z_i}$ where $x_i$, $y_i$, and $z_i$ are the three dimensions 
quoted in \citet{weaver2016}. The 1$\sigma$ errors in these radii are $\pm$2.5~km 
for Styx, Nix, and Kerberos and $\pm$8.5~km for Hydra.

Robust analyses of the HST imaging data yield satellite masses and 1$\sigma$ errors 
(in units of $10^{18}$~g) $m_S \lesssim$ 15 (Styx), $m_N = 45 \pm 40$ (Nix), 
$m_K = 16.5 \pm 9$ (Kerberos), and $m_H = 48 \pm 42$ (Hydra). Fits to the HST 
astrometry are rather insensitive to the masses of the small satellites 
\citep{brozovic2015}.  Using the adopted radii, the satellites have mass density 
$\rho_S \lesssim$ 25~\gcmc, $\rho_N = 1.3 \pm 1.3$~\gcmc, $\rho_K = 18 \pm 10$~\gcmc, 
and $\rho_H = 1.2 \pm 1.2$~\gcmc. Although the nominal densities for Nix and Hydra 
are reasonably close to the density of either Charon ($\rho_C$ = 1.70~\gcmc) or 
Pluto ($\rho_P$ = 1.85~\gcmc), results for Styx and Kerberos are unphysical.

For the calculations in this paper, we adopt the HST masses for Nix and Hydra.
Revising the mass density of Styx and Kerberos to the physically plausible
$\rho_S \approx \rho_K \approx$ 1~\gcmc\ yields masses, $m_S$ = 0.6 and $m_K$ = 1,
that are consistent with the detailed analyses of HST imaging data. Long-term
\nbody\ simulations with these masses yield stable systems over evolution times 
of 500~Myr. Test calculations show that doubling the adopted masses for Styx and
Kerberos does not change the results significantly.  We plan to investigate the 
masses of all of the satellites in more detail in a separate publication.

For massless tracers orbiting the \pc\ binary with semimajor axes 
$a_S \lesssim a \lesssim a_H$, plausible regions of stability are set by the
Hill radius
\begin{equation}
\rhill = \left ( { m \over { 3 ~ \mpc } } \right )^{1/3} ~ a,
\label{eq: rhill}
\end{equation}
where $m$ is the mass of a satellite, \mpc\ is the combined mass of Pluto and 
Charon, and $a$ is the semimajor axis of a nearby satellite.  The fifth column 
of Table~\ref{tab: sats} lists the Hill radius of each satellite. In systems 
with several massive planets or satellites, massive objects typically clear 
out a zone with a half-width $\delta a \approx K \rhill$ around their orbits, 
with $K$ = 8--10 \citep[e.g.,][and references therein]{wisdom1980,petit1986,
gladman1993,chambers1996,deck2013,fang2013,fabrycky2014,mahajan2014,pu2015,
morrison2016,obertas2017,weiss2018}.  

Naive application of these constraints leave little space for satellites with
$a_S \lesssim a \lesssim a_H$. Setting $K \approx$ 10 precludes other stable 
satellites between the orbits of Styx--Nix and Kerberos--Hydra, but allows moons 
in a small region $a \approx$ 54,000--56,000~km between the orbits of Nix and 
Kerberos. In terms of the binary separation \apc, this stable region has 
$a \approx 2.75 - 2.9 ~ \apc$.  The weaker constraint $K \approx$ 8 expands 
the stable region between Nix and Kerberos and enables a second stable region at 
$a \approx 2.22 - 2.27 ~ \apc$ (43,500--44,500~km) between the orbits of Styx and 
Nix.  Stable satellites between the orbits of Kerberos and Hydra are still prohibited.

The Hill condition also allows stable satellites with $a \approx 1.84 - 2.15 ~ \apc$
inside the orbit of Styx and $a \gtrsim 3.75 ~ \apc$ beyond the orbit of Hydra.  For 
$a \lesssim 2.15 ~ \apc$, it seems likely that the \pc\ binary, Nix, and Hydra will 
eventually drive out massive satellites. Far outside the orbit of Hydra 
($a \gtrsim 50 ~ \apc$), the gravity of Hydra, the Sun, and the major planets 
combine to preclude stable satellites for some range of $a$, $e$ and $i$ 
\citep{michaely2017}.  Inside this region, orbits are generally stable. 
Although some orbits with $a \approx 1.84 - 2.15 ~ \apc$ might be stable, 
we expect satellites with $a \gtrsim 3.75~\apc$ are stable on longer time scales. 

\subsection{Stability of Tracers in Circumbinary Orbits}
\label{sec: sims-stab-trc}

Although there have been many studies of the stability of circumbinary orbits
\citep[e.g.,][]{dvorak1989,holman1999,pilat2003,pichardo2005,musielak2005,
pichardo2008,verrier2009,farago2010,doolin2011,li2016}, only a few results are 
generally applicable to the \pc\ binary \citep[see also][and references 
therein]{youdin2012}. Nearly all orbits with $a \gtrsim$ 3~\apc\ are stable. 
For co-planar, prograde circumbinary satellites with inclination 
$\imath \approx$ 0\deg\ orbiting a \pc\ binary with 
$\epc \lesssim 5 \times 10^{-5}$ \citep{brozovic2015}, the innermost stable orbit 
has a semimajor axis $a_0 \approx$ 1.7--2 \apc. Some retrograde orbits 
($\imath \approx \pi$) with $a \approx$ 1--2 \apc\ are stable \citep{doolin2011}.  
Although orbits with $a \approx$ 1--3~\apc\ and $\imath \gg$ 0\deg\ are generally 
less stable than their $\imath \approx$ 0\deg\ counterparts, there are islands of 
stability with $a \approx$ 1--2~\apc\ and some $\imath$.

To identify plausible locations for small circumbinary satellites in the current 
\pc\ system, we follow sets of massless test particles orbiting the \pc\ binary.
In these initial tests, we do not include the four small satellites. Rather than
attempt to duplicate previous efforts in complete detail \citep[e.g.,][]{doolin2011}, 
our goal is to locate stable regions for prograde/retrograde orbits with small 
inclination to the orbital plane of the binary and for polar orbits with initial 
$\imath \approx$ 90\deg. To facilitate this goal, we define a `survivor fraction'
as the fraction of tracers in orbit after 1--2~Myr of evolution, $f_s = N_s / N_0$,
where $N_s$ is the number of survivors and $N_0$ is the initial number of tracers.
Table~\ref{tab: fracs} summarizes $f_s$ for these calculations.

For the first tests, we examine calculations of the orbital evolution of massless 
tracers with initial eccentricity $e_0 = 10^{-5}$ and a range of semimajor axes, 
$a_0$ = 1.0--1.5~\apc\ and $a_0$ = 1.45--2.1~\apc, on prograde ($\imath \approx$ 0), 
polar ($\imath \approx \pi/2$), and retrograde ($\imath \approx \pi$) orbits.  
Because we do not mix tracers with different inclinations, these tests require six 
distinct calculations with 3136 tracers in each calculation.  Within this set, 
only one tracer on polar orbits survives.  A comparison of the orbital evolution of 
this tracer with others on polar orbits suggests it will be ejected in $\lesssim$ 
1~Myr.  Thus $f_s$ = 0 for polar orbits with $a_0$ = 1.0--2.1~\apc. Among tracers 
on prograde orbits, none with $a_0$ = 1.0--1.5~\apc\ survive; however, 29\% of those 
with $a_0 \approx$ 1.45--2.1~\apc\ remain.  Tracers on retrograde orbits are more 
stable: 20\% (98\%) of the survivors have $a_0$ = 1.0--1.5~\apc\ (1.45--2.1~\apc).

\begin{figure}
\includegraphics[width=6.5in]{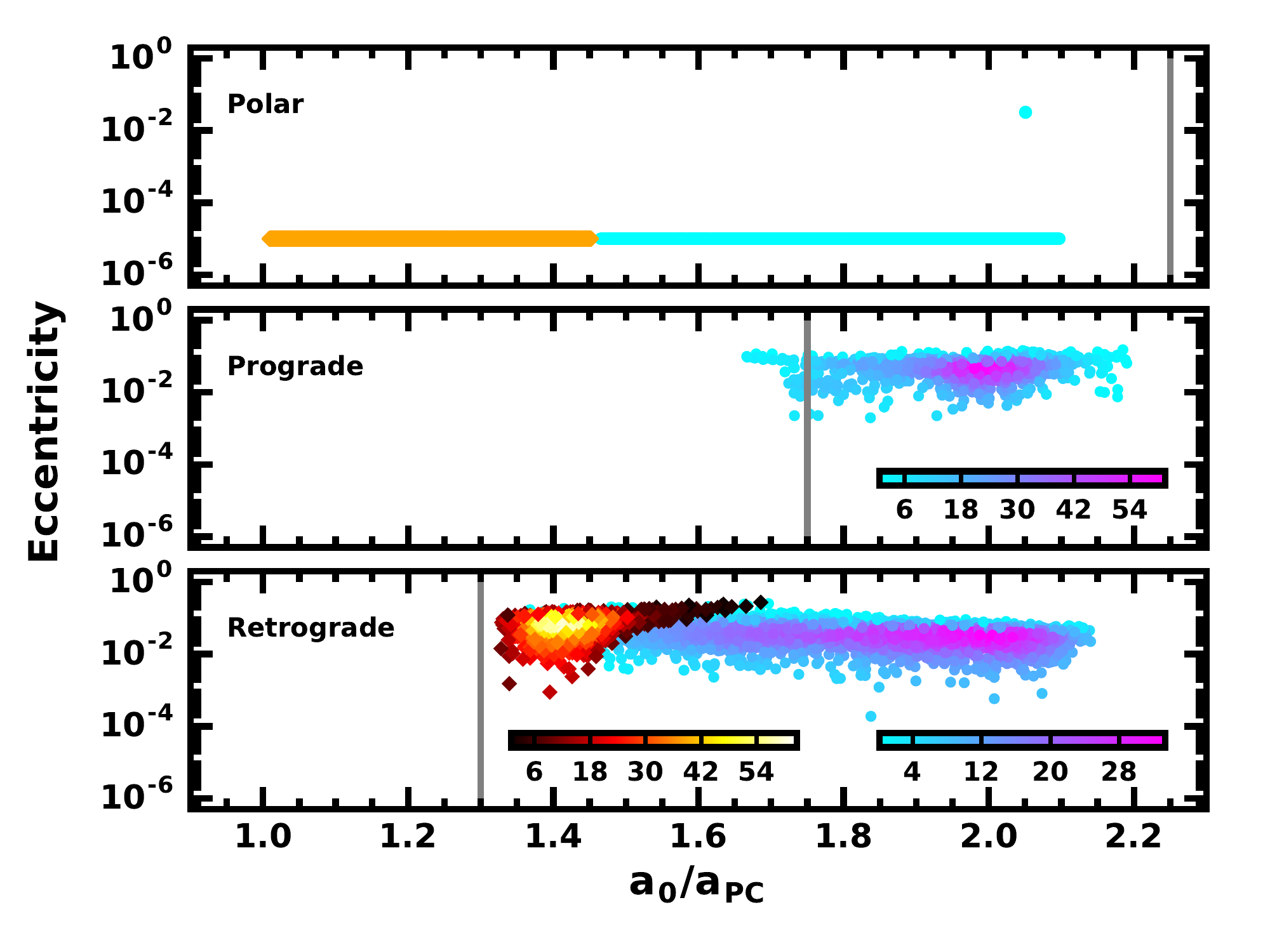}
\vskip 2ex
\caption{\label{fig: aebin1}
Distribution of semimajor axis $a$ and eccentricity $e$ for surviving massless 
test particles polar (upper panel), prograde (middle panel), or retrograde 
(lower panel) orbits around the \pc\ binary.  In the top panel, horizontal 
orange and cyan lines plot $(a_0, e_0)$ for each particle. Although we omit 
these data in other panels for clarity, other calculations have identical $e_0$.  
After 1--2~Myr of evolution, 
only one test particle with $(a,e) \approx$ $(2.05, 0.01)$ survives on a polar orbit.  
In the middle and lower panels, the colors of points within clouds at $e \approx$ 0.01 
indicate the density of survivors after 1--2~Myr, as indicated by the colorbars below
each cloud.  Among test particles with $a_0$ = 1.0--1.5~\apc\ ($a_0$ = 1.5--2.1~\apc), 
the color ranges from dark red (cyan) for low density to orange (light purple) for 
intermediate density to bright yellow (magenta) for high density.  Vertical grey lines 
indicate the approximate minimum $a$ for stable orbits from \citet{doolin2011}. 
Although there are no co-rotating survivors with $a_0 \approx$ 1.0--1.5~\apc\ in 
the middle panel, survivors with larger $a_0$ cluster at $a \approx$ 2~\apc\ and have 
a negligible density close to the expected minimum stable $a$. Retrograde survivors 
in the lower panel cluster at 1.4~\apc\ ($a_0 \approx$ 1.0--1.5~\apc) and at 
2.0~\apc\ ($a_0 \approx$ 1.5--2.1~\apc).  The minimum $a$ for both sets lies 
just outside the limit of \citet{doolin2011}.
}
\end{figure}

Fig.~\ref{fig: aebin1} shows the distribution of $a$ and $e$ for particles that 
survive for 1--2~Myr. The lone polar survivor has $a \approx$ 2.05~\apc\ and
$e \approx$ 0.04.  Among prograde tracers, survivors have $a \gtrsim$ 
1.70--1.75~\apc\ and $e \approx$ 0.01--0.1. Most are concentrated in a cloud 
with $a \approx$ 2~\apc\ and $e \approx$ 0.03. Retrograde survivors have
$a \gtrsim$ 1.3~\apc; the typical $e$ ranges from roughly 0.1 at 
$a \approx$ 1.4~\apc\ to 0.01 at $a \approx$ 2.1~\apc. 

\begin{figure}
\includegraphics[width=6.5in]{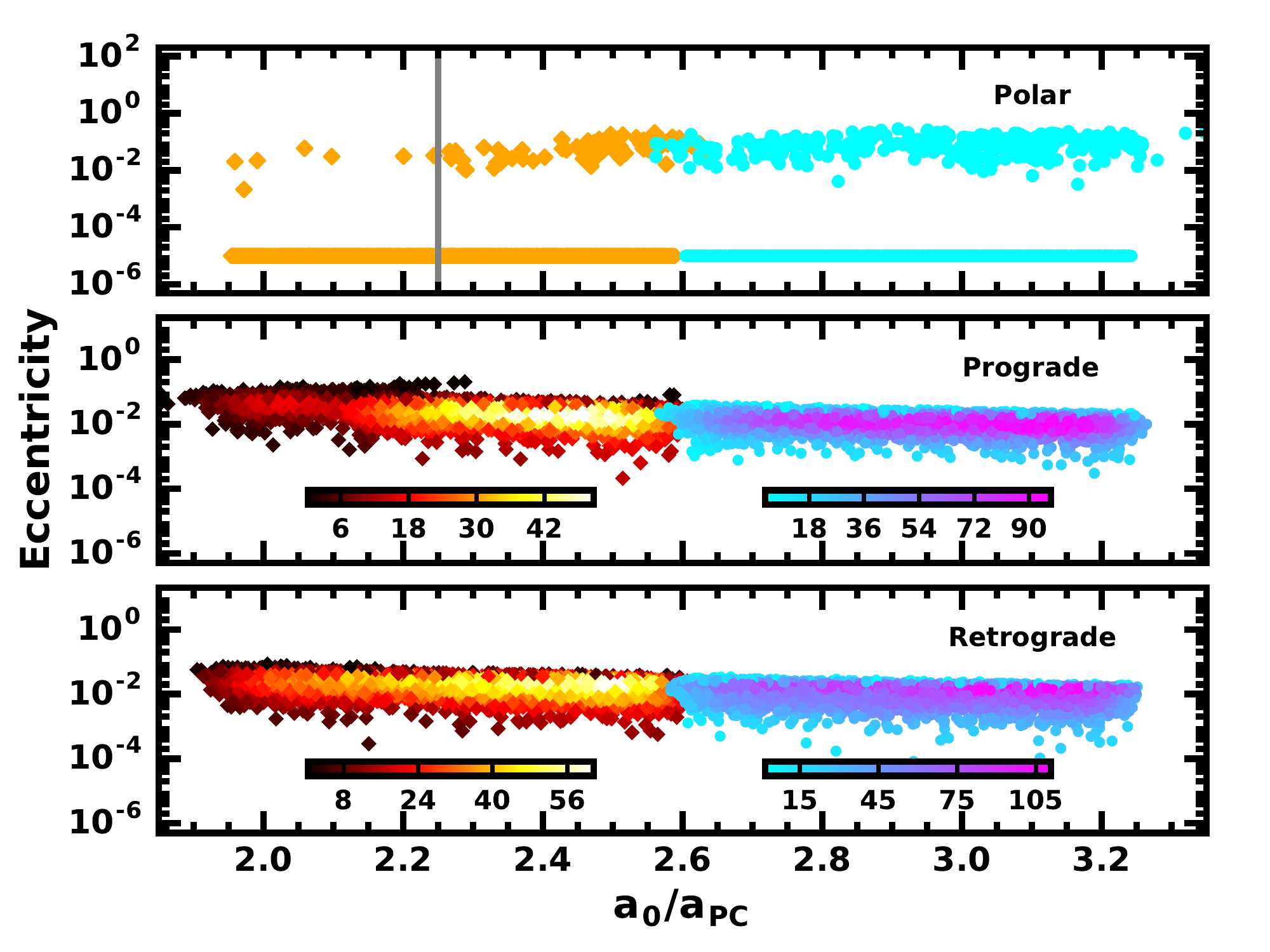}
\vskip 2ex
\caption{\label{fig: aebin2}
As in Fig.~\ref{fig: aebin1} for test particles with $a_0 \approx$ 2.0--3.2. 
Prograde and retrograde survivors lie in clouds where the typical $e$ declines
slowly with $a$. Few survivors on polar orbits lie inside the vertical grey 
line, which indicates the minimum stable $a$ of \citet{doolin2011}.
}
\end{figure}

Tracers with larger $a_0$ are more likely to survive. When $a_0$ = 
1.95--2.65~\apc\ ($a_0$ = 2.60--3.25~\apc), roughly 2\% (10\%) on polar orbits
survive for 1--2~Myr (Table~\ref{tab: fracs}). The survivor fraction is much
larger for tracers on prograde orbits -- 90\% for $a_0$ = 1.95--2.65~\apc\ and 
100\% for $a_0$ = 2.60--3.25~\apc.  All of the retrograde tracers survive, 

Fig.~\ref{fig: aebin2} shows $(a, e)$ for survivors with $a_0$ = 1.95--3.25~\apc. 
Prograde and retrograde tracers have a clear trend in $e(a)$, with 
$e \approx$ 0.01--0.1 at $a \approx$ 2~\apc\ and $e \approx$ 0.001--0.01 
at $a \approx$ 3.2~\apc. Sets of polar tracers have little or no trend in 
$e$ with $a$. However, there is an abrupt inner edge to the distribution of 
polar tracers at $a \approx$ 2.2~apc. For all sets of tracers, there is a 
broad range of $e$ at every stable $a$.

The minimum semimajor axes for circumbinary particles -- 
$a_c / \apc\ \approx$ 1.7 for prograde tracers, 2.2 for polar tracers, 
and 1.30--1.35 for retrograde tracers -- agree with previous results. 
For example, \citet{doolin2011} infer $a_c / \apc\ \approx$ 1.75 (prograde), 
2.2 (polar) and 1.3 (retrograde). The \citet{holman1999} fit to a suite of 
simulations yields $a_c / \apc\ \approx$ 2 for prograde orbits; retrograde 
orbits have a smaller $a_c$ \citep{weigert1997}.  The somewhat larger $a_c$ 
for prograde orbits from \citet{holman1999} matches the location of the high 
density cloud of survivors in Fig.~\ref{fig: aebin1}. Given the smaller number
of simulations with shorter duration performed by \citet{holman1999}, it is
possible that their calculations identified the most likely $a_c$ rather than 
the true $a_c$.

\subsection{Stability of Tracers in Circumbinary Orbits with Massive Satellites}
\label{sec: sims-trace}

We now consider the stability of tracers in systems with Pluto-Charon and the
four small satellites. The state vector of \citet{brozovic2015} establishes
initial positions and velocities of Pluto-Charon, Styx, Nix, Kerberos, and 
Hydra. For simplicity, we transform the coordinates to a cartesian system where 
the orbital plane of \pc\ defines $z$ = 0.
Tracers begin on nearly circular ($e_0 \approx 10^{-5}$) prograde orbits with 
low ($\imath \approx 10^{-5}$) or high ($\imath \approx \pi/2$) inclination. 
For three sets of calculations, the initial semimajor axis of each tracer is 
randomly distributed between $0.975 ~ a_i$ and $1.025 ~ a_{i+1}$ where $i$ is 
1 (Styx), 2 (Nix), 3 (Kerberos), and 4 (Hydra).  In a fourth (fifth) calculation, 
tracers have initial semimajor axes ranging from $0.7 ~ a_S$ to $1.025 ~ a_S$ 
($0.975 ~ a_H$ to $1.125 ~ a_H$).  The \nbody\ code follows the orbits of tracers 
until all have been ejected or for 10~Myr.

Among massless tracers on prograde orbits inside the orbit of Styx, roughly 20\%
survive 10~Myr of dynamical evolution. In the first year, nearly half are ejected;
less than a third remain after 0.5~Myr. At the end of the evolutionary sequence, 
the system ejects a few tracers per Myr. If this rate is maintained indefinitely,
all would be ejected in 100--200~Myr. It seems likely that the rate will slow;
thus, some are likely to survive for several Gyr.

The survival rate for tracers placed on prograde orbits outside the orbit of Styx 
and inside the orbit of Hydra is much smaller (Table \ref{tab: fracs}).  During 
the first Myr, nearly 90\% are ejected.  Over the next 9 Myr, the ejection rate 
slows considerably and eventually falls to roughly 1--2 tracers per Myr.  After 
10~Myr, the survival fractions are $f_s$ = 0.02 (Styx--Nix tracers), 0.004 
(Nix--Kerberos tracers), and 0.02 (Kerberos--Hydra tracers). Over the next 
100--200~Myr, the small but finite removal rate during the first 10~Myr implies 
the removal of all tracers initially placed inside the orbit of Hydra.

Tracers starting from just inside the orbit of Hydra to roughly $1.2 a_H$ have a 
larger survival rate. During the first 1--2~Myr of evolution, Nix and Hydra eject
nearly half of the tracers. After this initial flurry, the ejection rate slows 
to a trickle. At 10~Myr, $f_s$ = 0.45.

\begin{figure}
\includegraphics[width=6.5in]{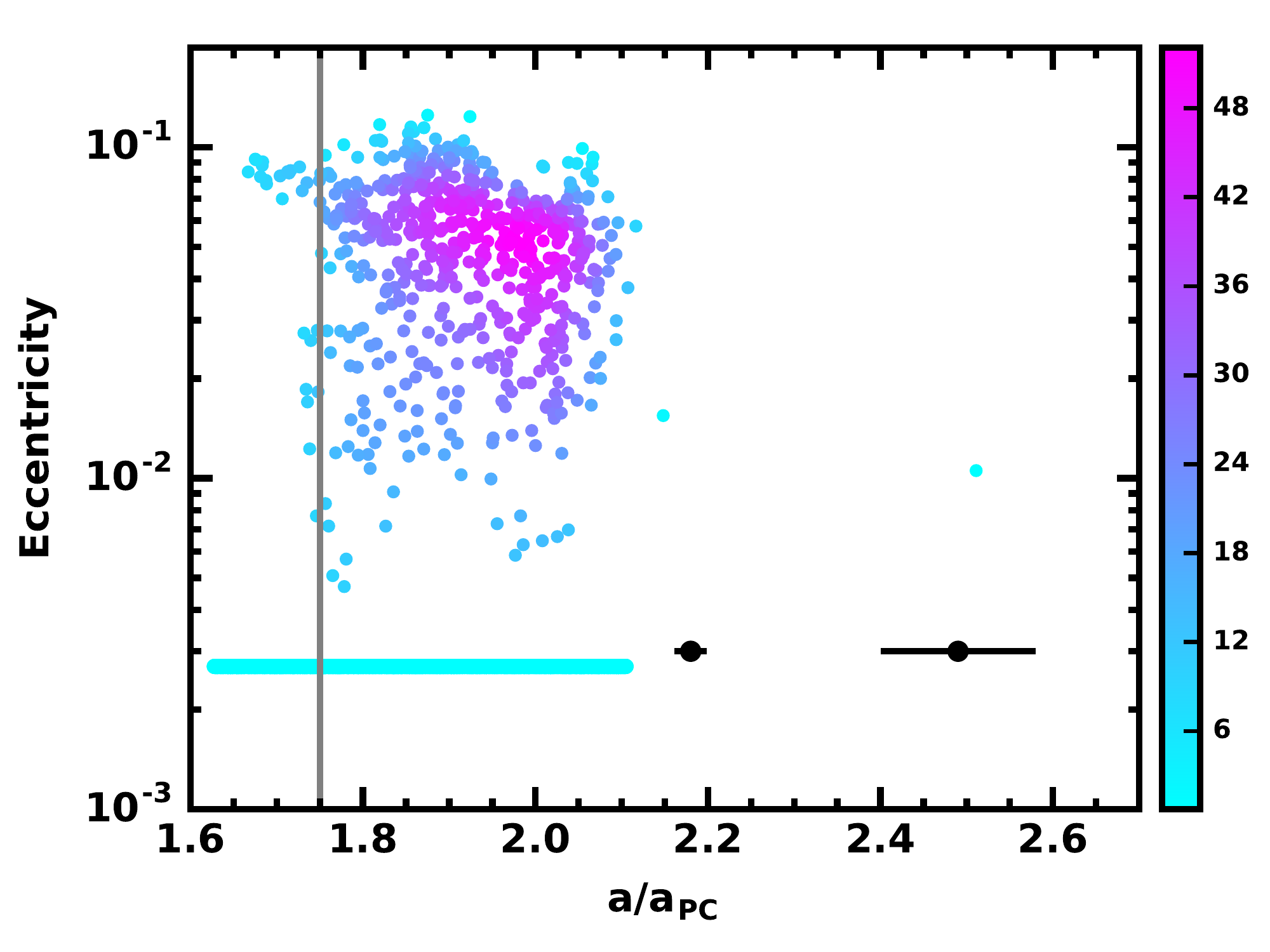}
\vskip 2ex
\caption{\label{fig: aepro1}
Distribution of $a$ and $e$ for surviving prograde test particles orbiting with 
the \pc\ satellites after 10~Myr. The vertical grey line indicates the position
of the innermost stable orbit from \citet{doolin2011}.  Black points plot 
the positions of the \pc\ satellites; horizontal lines extending from each 
point have half-width $\delta a = 2 \sqrt{3} \rhill$.  Cyan points at 
$e \approx 3 \times 10^{-3}$ indicate the initial range of $a$ for massless 
tracers co-rotating with the \pc\ binary. Smaller points represent the 
survivors after 10~Myr of dynamical evolution.  Colors of points within the cloud 
at $a \approx$ 1.8--2.1~\apc\ indicate the density of survivors, ranging from low 
(cyan) to intermediate (purple) to high (magenta) as indicated by the colorbar 
at the right of the plot.  Inside $a \approx$ 1.7--1.8~\apc, a handful of 
survivors will become unstable on time scales of $\lesssim$ 10~Myr.  Outside 
this limit, survivors are strongly clustered at $a \approx$ 1.9--2.0~\apc\ with 
a broad range of $e$.  
}
\end{figure}

\begin{figure}
\includegraphics[width=6.5in]{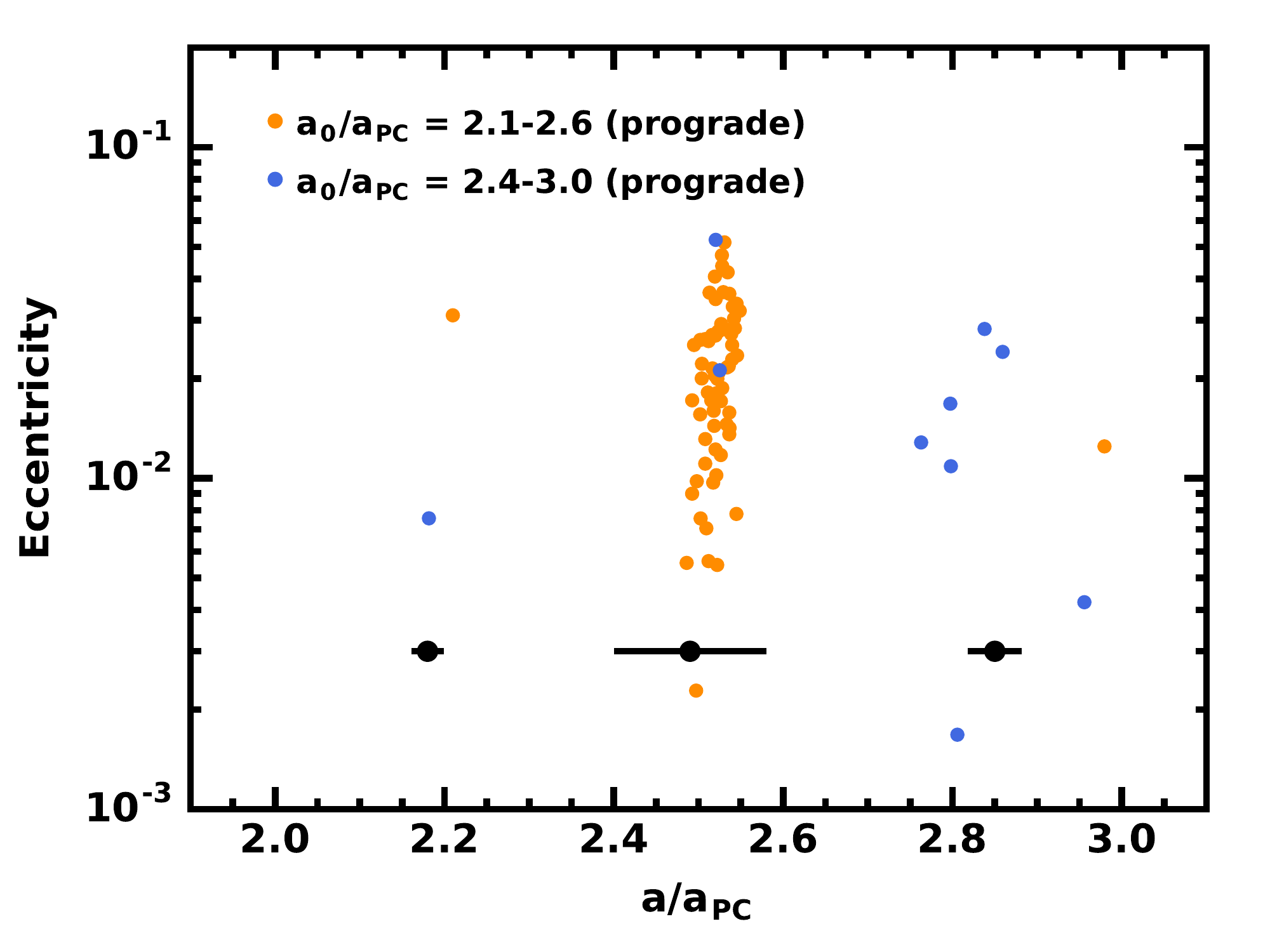}
\vskip 2ex
\caption{\label{fig: aepro2}
As in Fig.~\ref{fig: aepro1} for prograde test particles with 
$a_0$ = 2.1--2.6~\apc\ (orange points) or $a_0$ = 2.4--3.0~\apc\ (cyan points).
After 10~Myr, few test particles survive.
}
\end{figure}

\begin{figure}
\includegraphics[width=6.5in]{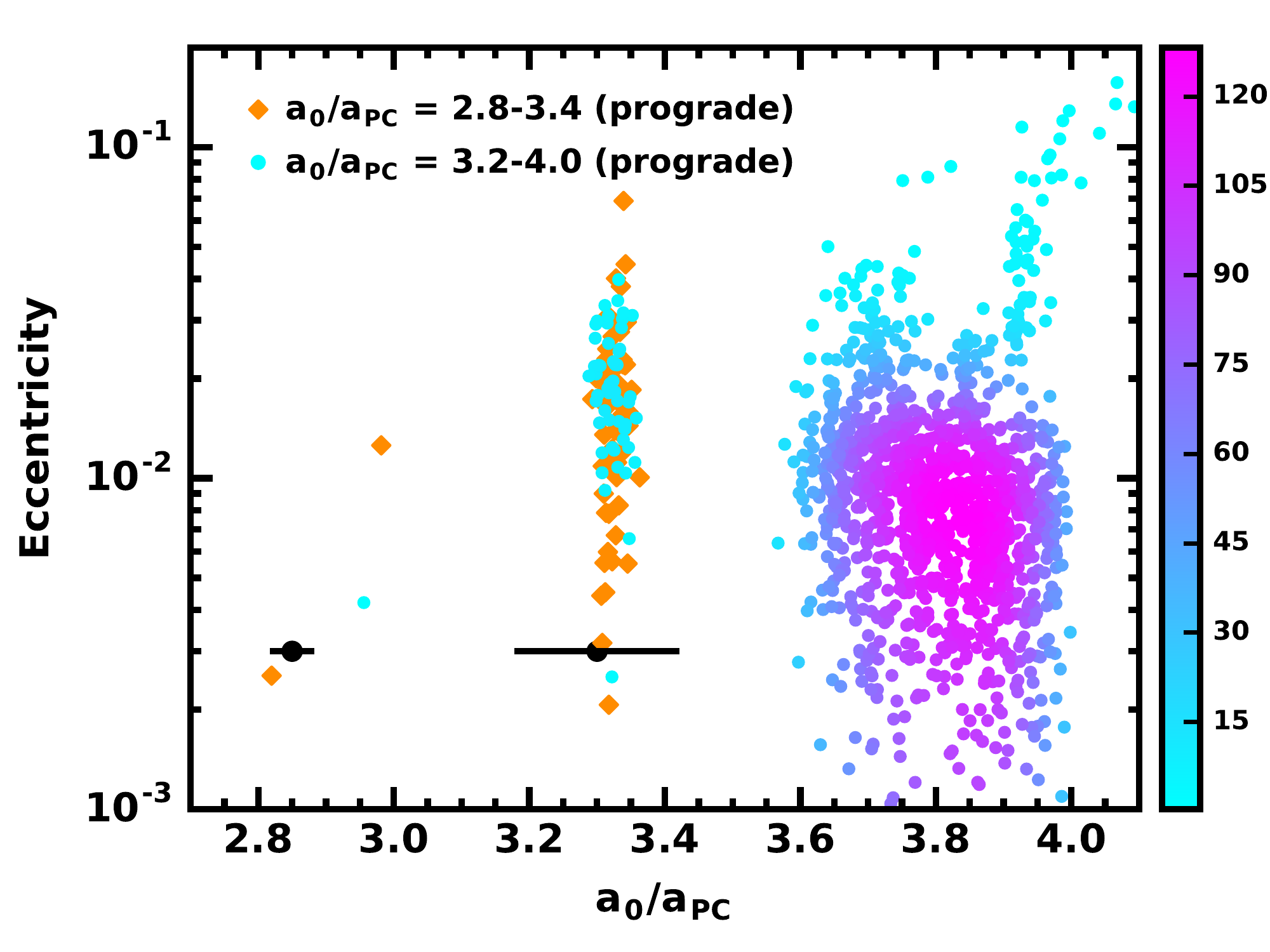}
\vskip 2ex
\caption{\label{fig: aepro3}
As in Fig.~\ref{fig: aepro1} for test particles with $a_0$ = 2.8--3.4~\apc\ (orange
points) or $a_0$ = 3.2--4.0~\apc\ (cyan points). Colors of points within the
cloud at $a \gtrsim$ 3.55~\apc\ indicate the density of survivors, ranging from
low (cyan) to purple (intermediate) to magenta (high) as indicated by the colorbar
to the right of the plot. Inside $a \approx$ 3.55~\apc, a handful of survivors 
will become unstable on time scales of $\lesssim$ 100~Myr. Thus, there are no 
stable orbits inside 3.55~\apc. Outside this limit, survivors are strongly clustered 
at $a \approx$ 3.8~\apc\ with a broad range of $e$. Within this group, the high $e$ 
objects likely become unstable on time scales of 100--200~Myr.
}
\end{figure}

For prograde tracers, the distributions of $(a, e)$ show several clear features.
Inside the orbit of Styx, the density of survivors peaks at 1.8--2.0~\apc\ 
(Fig.~\ref{fig: aepro1}).  Within the corotation zones of Nix and Hydra, a few tracers
orbit with $e$ = 0.005--0.05 (Fig.~\ref{fig: aepro2}--\ref{fig: aepro3}). There are 
essentially no tracers outside any of the corotation zones or in the corotation 
zones of Styx and Kerberos.  Survivors also have a strong concentration outside 
the orbit of Hydra, extending over a region bounded by $a \approx 3.6 - 4.0 ~ \apc$ 
and $e \approx 10^{-4} - 10^{-1}$, with a strong density maximum at 
$(a, e) \approx (3.8 ~ \apc, 0.005)$.  

Among tracers in the corotation zones of Nix and Hydra, the evolution of $e$ follows
a standard pattern. For several Myr, $e$ gradually grows from $\lesssim$ 0.001 to 0.05.
At some point, perturbations by the central binary, Nix, and Hydra generate 
$e \gtrsim$ 0.05 and the tracer begins to cross the orbit of either Nix or Hydra. 
After another few thousand years, the tracer leaves the system. 

Within the strong concentration of tracers inside the orbit of Styx and outside the 
orbit of Hydra, $e$ traces a similar evolution. Tracers that achieve $e \gtrsim$ 0.02 
suffer small oscillations in $a$ and an increasingly larger $e$ until the pericenter 
of their orbit approaches the orbit of Styx/Nix (tracers originally inside the orbit 
of Styx) or Hydra (tracers originally outside the orbit of Hydra). The gravity of
either Nix or Hydra then ejects them from the system. The upper envelope of the cyan 
points in the upper right corner of Fig.~\ref{fig: aepro3} show the clear growth of $e$ 
with increasing $a$ that characterizes dynamical ejections \citep[see also][]{dunc1997,
gladman2002}.

The evolution of polar tracers is somewhat different. Unlike prograde tracers, polar 
tracers with orbits that cross within the corotation zone of a massive satellite are 
rapidly ejected from the system. Tracers with orbits that cross between the satellites
are harder to remove. These tracers spend most of their time well outside the 
`clearing zone' of the massive satellites and take somewhat longer to eject. Still, 
some regions are cleared rapidly: it takes only 50~kyr to eject 100\% (90\%) of polar 
tracers with initial semimajor axes inside the orbit of Styx (between the orbits of 
Styx and Nix). Despite the slow removal rate for tracers inside the orbit of Nix after
0.1~Myr, all are lost after 6~Myr. 

Polar tracers survive more easily outside the orbit of Nix (Table~\ref{tab: fracs}). 
For orbits between Nix--Kerberos (Kerberos--Hydra), it takes 0.1~Myr (0.3~Myr) to 
reduce the initial number of tracers by 50\%. The removal rate then slows considerably. 
After 10~Myr, the survivor fractions are $f_s$ = 0.21 (Nix--Kerberos tracers) and 
$f_s$ = 0.14 (Kerberos--Hydra tracers). To examine the removal rate at later times, 
we extended these calculations to 20~Myr. At this point the survivor fractions have 
dropped to $f_s$ = 0.18 (Nix--Kerberos) and $f_s$ = 0.08 (Kerberos--Hydra). In both 
cases, the removal rate at 20~Myr suggests that the \pc\ satellite system will 
eventually eject all of the tracers with polar orbits between Nix and Hydra, on a 
time scale of $\sim$ 50~Myr for Nix--Kerberos tracers and $\sim$ 30~Myr for 
Kerberos--Hydra tracers. 

The longer lifetime for polar tracers between the orbits of Nix and Kerberos is 
due to the larger Hill spacing factor ($K$ = 16) relative to the $K$ = 10 factor 
for the Kerberos--Hydra pair.  With Hydra's larger nominal mass and its closer 
orbit to Kerberos, the volume available for extra satellites is much larger 
between Nix and Kerberos than it is between Kerberos and Hydra. The larger (smaller)
volume results in a slower (faster) removal process for tracers on unstable 
orbits between Nix and Kerberos (Kerberos and Hdyra).

Outside the orbit of Hydra, tracers on polar orbits are much more stable. 
After 1~Myr (3~Myr), the massive satellites have ejected only 30\% (36\%) of 
tracers on polar orbits with $a_0$ = 3.2--4.0~\apc. After 20~Myr, the survivor
fraction exceeds 0.5 (Table~\ref{tab: fracs}). Based on the slow rate of removal
for this set of polar tracers, many will survive for $\gtrsim$ 500~Myr.

\begin{figure}
\includegraphics[width=6.5in]{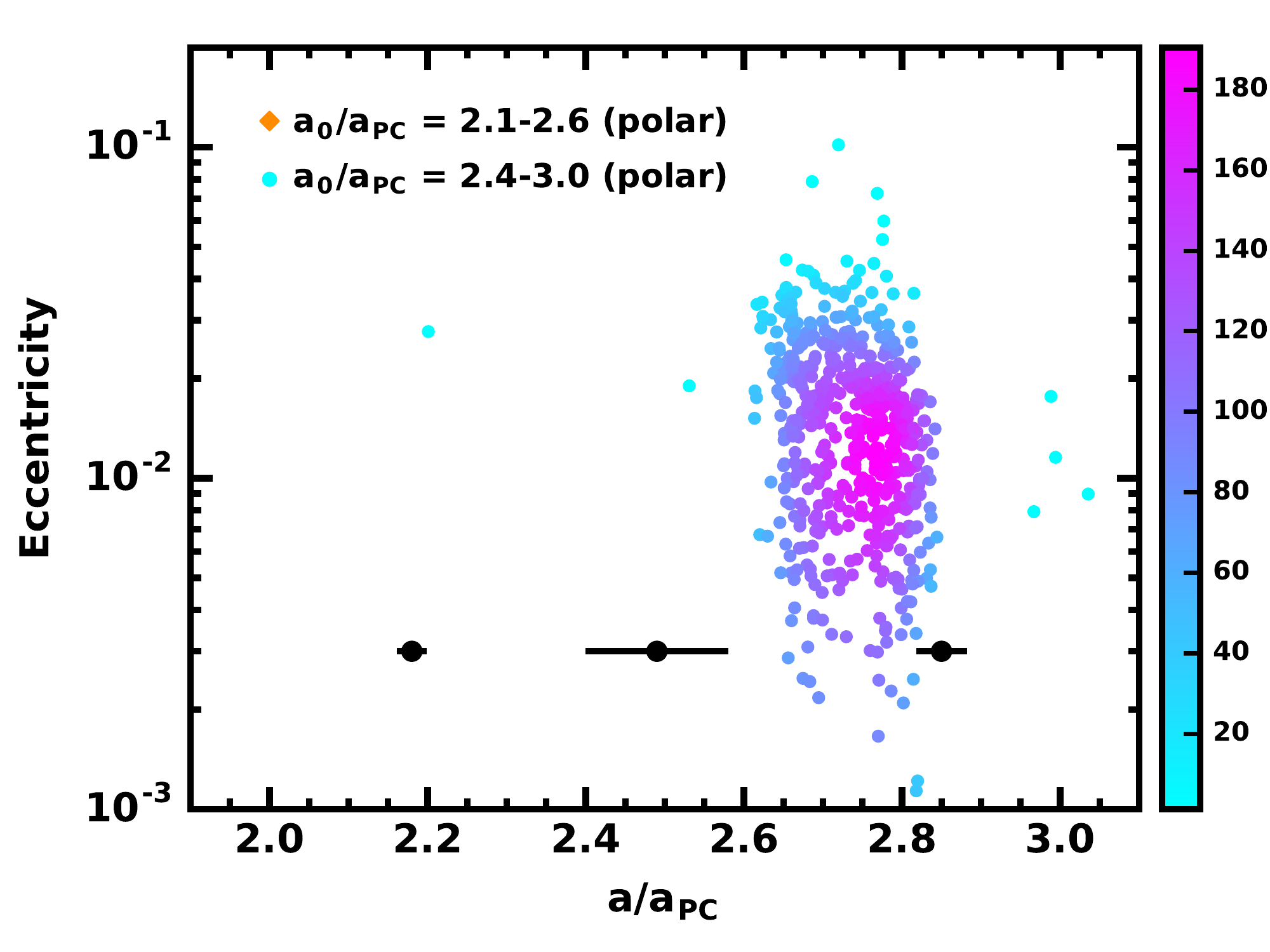}
\vskip 2ex
\caption{\label{fig: aepol1}
As in Fig.~\ref{fig: aepro2} for test particles on polar orbits. The few survivors
remaining after 10 Myr of evolution are likely to be ejected after 100--200~Myr.
}
\end{figure}

\begin{figure}
\includegraphics[width=6.5in]{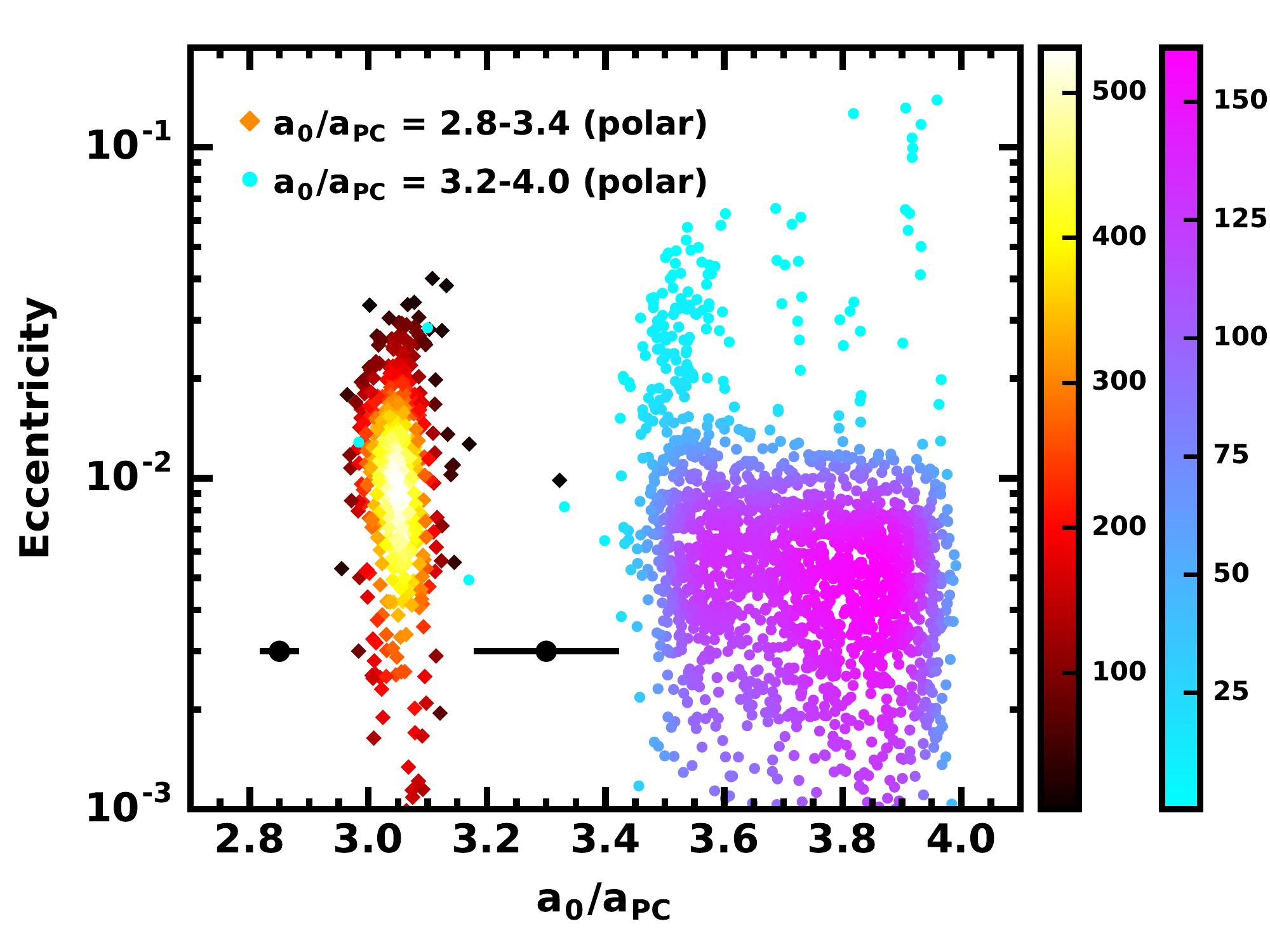}
\vskip 2ex
\caption{\label{fig: aepol2}
As in Fig.~\ref{fig: aepol1} for particles with larger $a_0$.
}
\end{figure}

Despite differences in the dynamical evolution between tracers on polar orbits and 
tracers on low inclination prograde orbits, the distributions of $(a,e)$ have some
common features (Figs.~\ref{fig: aepol1}--\ref{fig: aepol2}). At any time, the range 
in $e$ for survivors is very large with a maximum $e$ of roughly 0.05 for prograde 
orbits and roughly 0.01 for polar orbits. Tracers excited to larger $e$ are rapidly 
ejected. In both sets of calculations, dynamical evolution generates a dense cloud 
of tracers with $(a, e) \approx (3.8~\apc, 0.05)$. Within this cloud, the maximum 
stable $e$ is 0.01--0.02 instead of 0.05--0.10.

\begin{figure}
\includegraphics[width=6.5in]{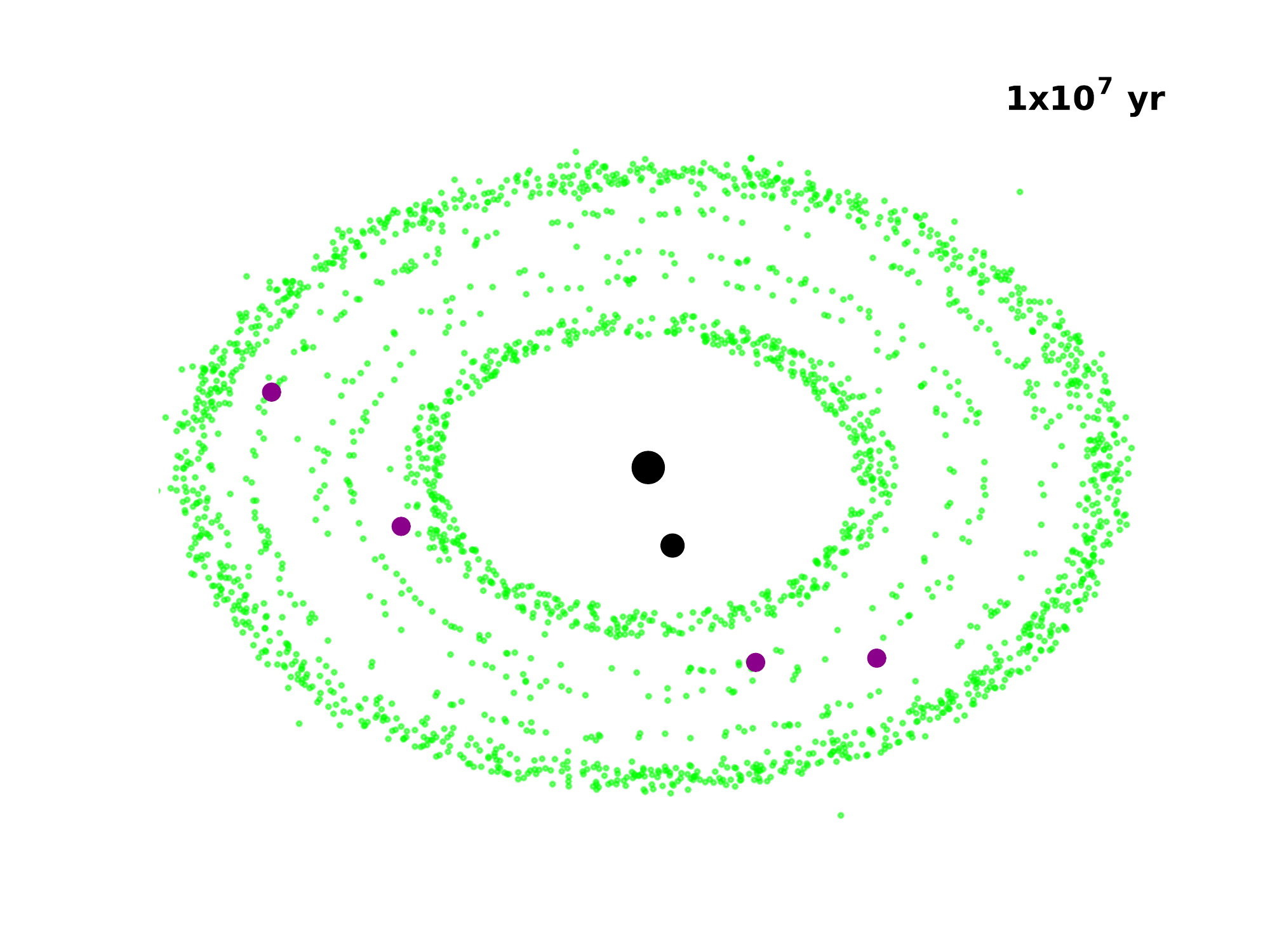}
\vskip 2ex
\caption{\label{fig: xypro}
Positions of Pluto-Charon (large black dots), the four small satellites (purple dots),
and surviving prograde, massless tracers (small green dots) after 10~Myr of dynamical 
evolution. The system is viewed at an inclination angle of 45\deg\ relative to the 
orbital plane.  Most survivors lie just inside the orbit of Styx or outside the orbit 
of Hydra.  Several tracers orbit within the co-rotation zones of Nix or Hydra. A 
time-lapse animation illustrates the loss of massless tracers with time.
}
\end{figure}

\begin{figure}
\includegraphics[width=6.5in]{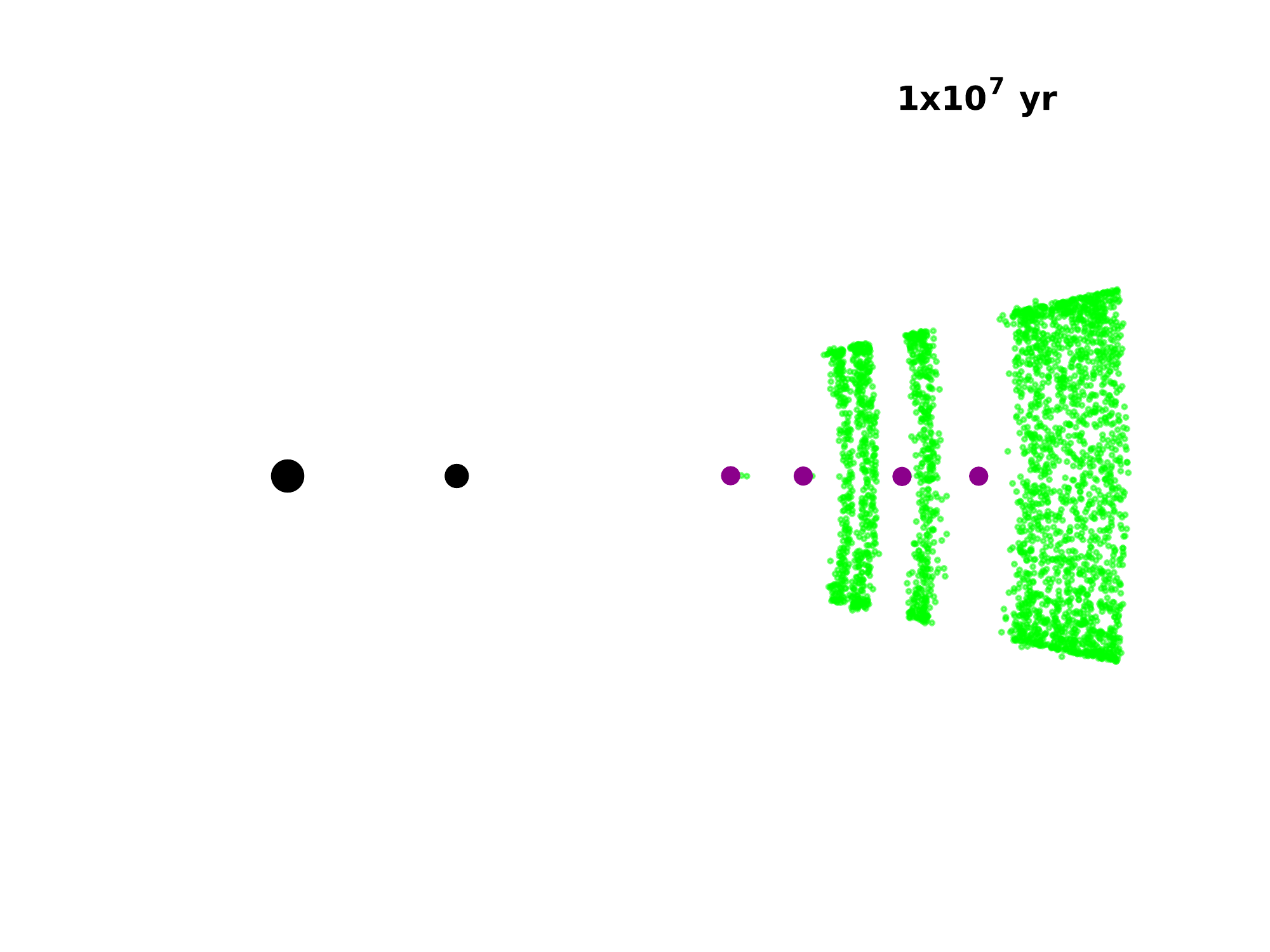}
\vskip 2ex
\caption{\label{fig: xypol}
As in Fig.~\ref{fig: xypro} for tracers on polar orbits. The view is in the 
orbital plane, plotting the semimajor axis $a$ on the $x$-axis and the $z$
distance from the orbital plane on the $y$-axis.  Most survivors lie just 
outside the orbit of Hydra.  Some tracers survive between the orbits of 
Nix--Kerberos and the orbits of Kerberos-Hydra.  Tracers on polar orbits do 
not survive inside the orbit of Nix.  A time-lapse animation illustrates 
the loss of massless tracers with time.
}
\end{figure}

To conclude this section, Figs.~\ref{fig: xypro} and \ref{fig: xypol} display the
configuration of tracers within the \pc\ satellite system after 10~Myr.  Animations 
associated with each figure illustrate the time evolution of the complete population.
Because we calculate the evolution of tracers in bands that overlap each satellite,
the overall population of tracers is somewhat larger along the orbit of each satellite
than in the volume between the satellites. 

For the ensemble of prograde tracers, it takes 100--300~yr for Nix and Hydra to begin
clearing out particles along their orbits. Slight density maxima appear along the orbits
of the lower mass satellites. Over roughly 3000~yr, Nix nearly clears out material in
its Hill volume, except for a narrow co-rotation zone. With a longer orbital period,
Hydra clears out a similar region in 10--30~kyr. During this period, the satellites
eject tracers in the density maxima along the orbits of Styx and Kerberos. As the 
evolution proceeds, the satellites clear out the Hill volumes of Styx and Kerberos 
(including their co-rotation zones), while Nix and Hydra continue to work on removing 
tracers inside their co-rotation zones (0.1--0.3~Myr).  Once the evolutionary sequence 
is complete, there are large sets of tracers just inside the orbit of Styx and just 
outside the orbit of Hydra, along with a few tracers in the co-rotation zones of Nix 
and Hydra. Otherwise, the system is fairly empty (Fig.~\ref{fig: xypro}).

The evolution of polar tracers shows several contrasting features (Fig.~\ref{fig: xypol} 
and associated animation). Without the four circumbinary satellites, polar tracers 
inside the orbit of Styx are unstable (Fig.~\ref{fig: aebin1}). Nearly all of these 
tracers disappear within 1000~yr.  As the binary ejects these tracers, Nix and then 
Hydra begin to clear polar tracers that intersect their orbits. Unlike systems of 
prograde tracers in the orbital plane of the binary, polar tracers orbiting in the 
co-rotation zones of each satellite are ejected as rapidly as other tracers within 
the Hill volume. After 10~kyr (100~kyr), the Hill volume of Nix (Hydra) is nearly empty.
As Nix completes clearing its Hill volume, the central binary ejects the last few 
tracers inside the orbit of Styx; Kerberos also begins to eliminate tracers from its 
orbit. At 1~Myr, the volume inside the orbit of Nix and the Hill volumes of Kerberos 
and Hydra are nearly devoid of tracers. After 10--20~Myr, many tracers remain between 
the orbits of Nix--Kerberos and Kerberos--Hydra and well outside the orbit of Hydra.

As the \pc\ binary and small satellites clear polar tracers out of the corotation zones
and the region inside the orbit of Nix, they also clear out material near the 9:2 
resonance with the orbit of the central binary. Visible as a narrow dark band in the 
positions of tracers between the orbits of Nix and Kerberos in Fig.~\ref{fig: xypol}, the 
density of tracers in this region is roughly 60\% of the density in the rest of this group.
The outer edge of the distribution of tracers between the orbits of Kerberos and Hydra is 
close to the 11:2 resonance. However, there are no tracers between this resonance and the
orbit of Hydra. Outside the orbit of Hydra, there is another drop in density at the 7:1 
resonance with the binary. This drop is not visible as a dark band in Fig.~\ref{fig: xypol};
yet, the density of particles is roughly 70\% of the density in the rest of the group outside
the orbit of Hydra. Although our calculations did not examine the evolution of polar tracers
with more distant orbits, the central binary and the small satellites remove polar tracers at
the n:1 resonances (with n = 2--7) or the n:2 resonances (with n = 3, 5, 7, 9, 11, and 13) on
time scales of $\lesssim$ 1--10~Myr.

\subsection{Stability of Low Mass Satellites}
\label{sec: sims-outer}

Although the calculations with massless tracers establish likely locations for 
small particles in the \pc\ system, they do not allow us to constrain sets of
stable orbits for small satellites with mass. To test whether massive satellites 
can have as long lifetimes as massless tracers, we select a group of massless, 
prograde test particles that have survived 10~Myr of circumbinary evolution 
(Figs.~\ref{fig: aepro1}--\ref{fig: aepro3}), assign each a radius of 2~km and 
a mass of $4 \times 10^{16}$~g ($\rho_s$ = 1.2~\gcmc), and evolve them with \pc, 
Styx, Nix, Kerberos, and Hydra. To avoid gravitational interactions among the 
survivors, we follow the evolution of only one survivor in addition to the 
central binary and the four known satellites of \pc.  

\begin{figure}
\includegraphics[width=6.5in]{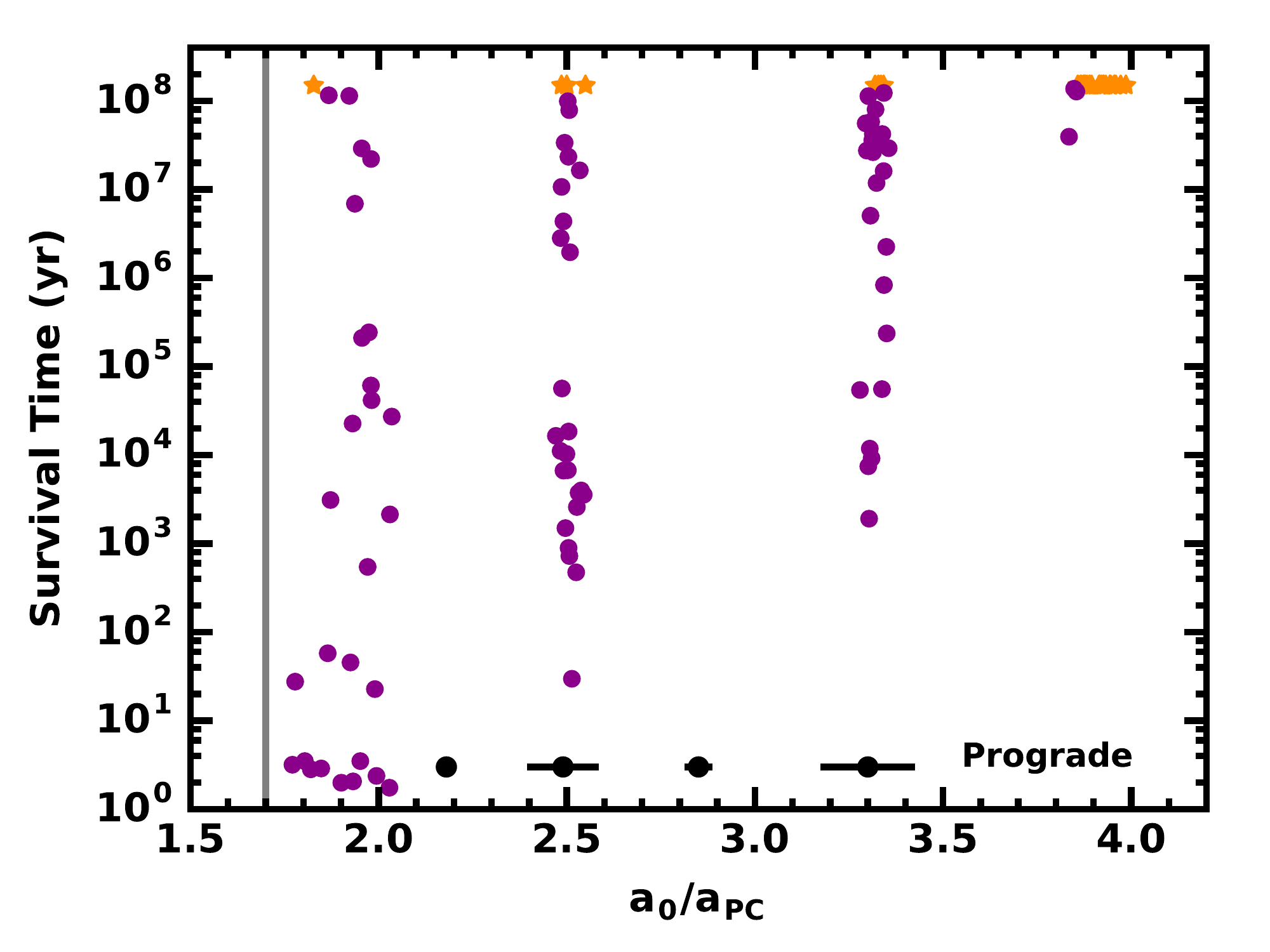}
\vskip 2ex
\caption{\label{fig: pro-moons}
Survival time of massive satellites ($r$ = 2 km; $m = 4 \times 10^{16}$~g)
as a function of initial semimajor axis. Purple points: ejected satellites;
orange stars: upper limits.  Initial orbits are selected from the survivors 
of test particles placed in nearly circular orbits with 
$a_0$ = 2.1--2.6~\apc\ (Fig.~\ref{fig: aepro1}, orange points),
$a_0$ = 2.8--3.4~\apc\ (Fig.~\ref{fig: aepro2}, orange points), and
$a_0$ = 3.2--4.0~\apc\ (Fig.~\ref{fig: aepro2}, magenta points). 
}
\end{figure}

For each set of survivors, the lifetime as a massive satellite is a strong function
of initial conditions (Table~\ref{tab: fracs}; Fig.~\ref{fig: pro-moons}).  Among 
those with orbits inside the orbit of Styx (near the corotation zone of Nix), 
survival times range from 1~yr to $\sim$ 100~Myr (30~yr to $\gtrsim$ 100~Myr).  
Satellites starting within the corotation zone of Hydra last somewhat longer, 
from roughly 20~kyr to $\gtrsim$ 100~Myr.  Despite the large range in lifetimes, 
90\% of the satellites initially inside the orbit of Styx and near the corotation 
zone of Nix are ejected within 100~Myr; 79\% of satellites initially near the 
corotation zone of Hydra are ejected after 100~Myr.  To examine the survival rate 
at later times, we extended calculations for survivors to 150~Myr.  At this point, 
the survivor fractions are $f_s$ = 0.10 (satellites inside the orbit of Styx), 
0.10 (satellites co-rotating with Nix), and 0.14 (satellites co-rotating with Hydra). 
Based on the gradual increase in $e$ and $\imath$ for these survivors at 100--150~Myr,
the likelihood that 2~km satellites inside the orbit of Hydra survive for the age 
of the solar system is small.

Massive moons on prograde orbits outside the orbit of Hydra are stable. After 
100~Myr of circumbinary evolution, $\sim$ 3.5\% of the satellites are ejected. 
Another 7\% are ejected after 150~Myr.  For each surviving satellite, there are 
no obvious trends in the evolution of $a$, $e$, or $\imath$.  Although it is 
possible that the orbits of some satellites might suffer significant perturbations 
at later times, the long lifetimes of these moons suggests that many are stable 
on 4--5~Gyr time scales.

To investigate the potential for small moons on polar orbits, we select a group 
of tracers that survive 10--20~Myr of evolution on polar orbits among the known 
\pc\ satellites (Figs.~\ref{fig: aepol1}--\ref{fig: aepol2}). As with prograde 
moons, we assign each a radius of 2~km and evolve each one together with the known 
satellites. Because polar tracers do not survive inside the orbit of Nix, we
only consider the evolution of small moons between the orbits of Nix--Kerberos
and Kerberos--Hydra and outside the orbit of Hydra.

\begin{figure}
\includegraphics[width=6.5in]{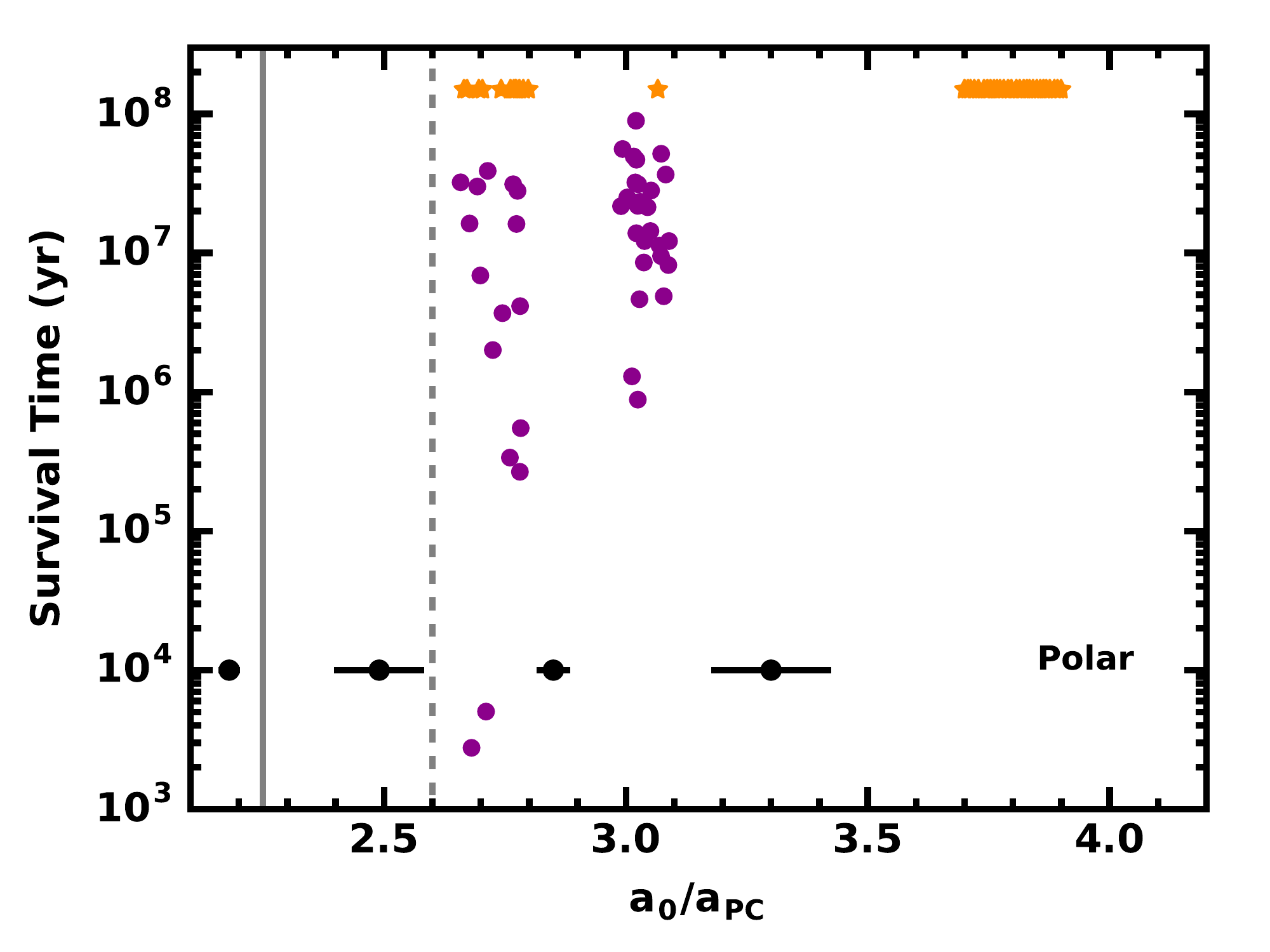}
\vskip 2ex
\caption{\label{fig: pol-moons}
As in Fig.~\ref{fig: pro-moons} for massive satellites ($r$ = 2 km; 
$m = 4 \times 10^{16}$~g) on polar, circumbinary orbits.  Initial orbits 
are selected from the survivors of test particles placed in nearly circular 
orbits with $a_0$ = 2.6--2.85~\apc\ (Fig.~\ref{fig: aepol1}, orange points),
$a_0$ = 2.95--3.1~\apc\ (Fig.~\ref{fig: aepol2}, orange points), and
$a_0$ = 3.7--3.9~\apc\ (Fig.~\ref{fig: aepol2}, magenta points). 
}
\end{figure}

Between the orbits of Nix--Kerberos and Kerberos--Hydra, several small moons survive
100~Myr of dynamical evolution (Table~\ref{tab: fracs}; Fig.~\ref{fig: pol-moons}). 
Moons orbiting between Nix and Kerberos often have short lifetimes of 1--10~kyr; 
others survive for less than 1~Myr. Roughly 21\% complete a 100~Myr calculation on 
a stable orbit. With a minimum lifetime of roughly 1~Myr, moons on polar orbits 
between Kerberos and Hydra generally last somewhat longer. However, few survive 
for 10~Myr. After 100~Myr, only $\sim$ 3.5\% are still on fairly stable orbits.

Outside the orbit of Hydra, small moons on polar orbits generally survive for 100~Myr.
For these satellites, perturbations from the known moons are fairly small. Orbiting
well inside the \pc\ Hill sphere, these satellites are also fairly immune to jostling
from the giant planets and other passersby. As with small moons on prograde orbits 
outside the orbit of Hydra, we expect that many of these can survive for $\gtrsim$ 1~Gyr.

\subsection{Summary}
\label{sec: sims-summ}

Direct \nbody\ simulations of circumbinary satellites confirm previous conclusions
for the semimajor axis of the innermost stable orbit \citep[e.g.,][]{doolin2011}. 
Massless prograde (retrograde) satellites with small $\imath$ relative to the plane
of the binary orbit are unstable when the initial semimajor axis 
$a_0 \lesssim$ 2.1~\apc\ ($a_0 \lesssim$ 1.7~\apc). Stable satellites on polar 
orbits must have $a_0 \gtrsim$ 2.2~\apc.

For the adopted masses of Styx, Nix, Kerberos, and Hydra, possible orbits for other 
stable satellites are tightly constrained. On time scales ranging from a few yr to 
10~Myr, nearly all massless tracers with prograde or polar orbits and 
$0.95~a_S \lesssim a \lesssim 1.1 a_H$ are ejected. Survivors on prograde orbits lie 
well inside the orbit of Styx or within the corotation zones of Nix and Hydra. Despite
the lack of polar survivors inside the orbit of Nix, some tracers remain on orbits
between Nix--Kerberos or Kerberos--Hydra.  Based on the time evolution of $e$ and 
the loss rate at 5--20~Myr, we suspect nearly all will be ejected within 100~Myr. 
For prograde and polar orbits, many tracers outside the orbit of Hydra 
($a \approx$ 3.6--4.0~\apc) are stable on $\gtrsim$ 10~Myr time scales. The orbital 
evolution of these tracers suggests they will remain stable over Gyr time scales.

Experiments with massive satellites confirm these conclusions.  Moons with 
radius $r$ = 2~km and mass $m = 4 \times 10^{16}$~g on prograde orbits with 
initial semimajor axis $a_0 \approx$ 1.7--2.1~\apc\ are unstable on time scales 
ranging from a few yr to 100--150~Myr. Prograde satellites corotating in the orbits 
of Nix or Hydra are ejected on time scales of 30~yr to 150~Myr. Small moons outside 
the orbit of Hydra ($a \approx$ 75,000~km to 80,000~km) survive for 150~Myr and 
are likely on stable orbits.

Small moons on polar orbits also survive 100~Myr of dynamical evolution. Among 
those with $a_0$ between the orbits of Nix--Kerberos (Kerberos--Hydra), a few 
remain on stable orbits. However, nearly all are ejected. Satellites with polar 
orbits beyond Hydra are generally stable.

These results agree with expectations based on the Hill radius of each satellite.
From previous calculations of multi-planet or multi-satellite systems orbiting 
single or binary central objects, stability requires $K \gtrsim$ 8--10 
\citep[e.g.,][]{wisdom1980,petit1986,gladman1993,chambers1996,deck2013,fang2013,
fabrycky2014,mahajan2014,pu2015,morrison2016,obertas2017}. With no stable tracers
between the orbits of Styx and Hydra, our calculations support the more 
conservative $K \gtrsim$ 10. 

The conservative limit for satellite stability disagrees with results from 
\citet{porter2015}, who predicted stable locations for satellites prior to the
\nh\ flyby. After re-analyzing HST data and suggesting a somewhat smaller
(larger) mass for Nix (Hydra), they describe a suite of numerical calculations
which indicate a broad range of stable orbits for particles between the orbits 
of each pair of satellites. Formally, these results imply $K \lesssim$ 2--3.
With typical durations of 1700~yr, however, their integrations are too short to 
infer robust lifetimes for extra satellites in the \pc\ system 
\citep[e.g.,][]{youdin2012}. Although \orch\ calculations confirm a large survivor
fraction for tracers with $a_S \lesssim a_0 \lesssim a_H$ at 1000--3000~yr 
(see the animations associated with Figs.~\ref{fig: xypro}--\ref{fig: xypol}), 
longer-term calculations demonstrate that particles with 
$0.95~a_S \lesssim a_0 \lesssim 1.1~a_H$ are ejected on a broad range of time 
scales that are often much larger than a few thousand years 
\citep[see also][]{youdin2012}.  Based on these longer integrations, we conclude 
that stable orbits with $a_S \lesssim a \lesssim a_H$ for additional satellites
are rare.

%The long lifetimes of massless tracers in our calculations are consistent with 
%results from \citet{youdin2012}, which considered the stability of a massless 
%Kerberos in a system with \pc, Nix, and Hydra. Using a 15th-order Radau integrator
%\citep{everhart1985} in the Swifter package\footnote{Publicly available at 
%http://www.boulder.swri.edu/swifter/}, they infer lifetimes ranging from 1~Myr to
%$\gtrsim$ 100~Myr for the nominal masses of Nix and Hydra adopted here. Although
%the masses and orbital configuration of the four small satellites considered here 
%differs considerably from those examined in \citet{youdin2012}, the similarity of
%lifetimes in the two sets of calculations is reassuring.

\section{OBSERVATIONAL PROGRAMS}
\label{sec: obs}

To construct observational programs to detect new satellites and to improve constraints
on the known \pc\ satellites, we adopt the nominal satellite properties in 
Table~\ref{tab: sats}.  At a distance from the Earth of roughly 40 AU ($6 \times 10^{14}$~cm), 
Hydra is 2\arcsec\ away from Pluto. Styx is somewhat closer, at an angular distance of 
1\secpoint3.  The circumference of Hydra's (Styx's) orbit is 12\secpoint5 
(8\arcsec); during its 38~d (20~d) orbital period, Hydra (Styx) moves an arcsec 
every 3.33~d (2.52~d), equivalent to 0.01 (0.0075) arcsec per hr.  From the 
standpoint of telescopic observations over several hours, Hydra, Styx, and the 
other satellites are effectively stationary with respect to Pluto. 

\subsection{Direct Imaging with JWST}
\label{sec: obs-jwst}

To explore options for detecting \pc\ satellites with JWST, we rely on 
published descriptions of the instruments \citep[e.g.,][]{beichman2012,
doyon2012,rieke2015} and sensitivity estimates from the {\it JWST 
Pocket Guide}\footnote{e.g., the `Pocket Guide' at 
https://jwst.stsci.edu/instrumentation/miri}.  With fields-of-view ranging 
from 74\arcsec\ $\times$~113\arcsec\ (MIRI; the Mid-Infrared Instrument) 
to 2\minpoint2~$\times$~2\minpoint2 (NIRISS; Near-Infrared Imager and Slitless 
Spectrograph) to 2 $\times$ 2\minpoint2~$\times$~2\minpoint2 (NIRCam; 
Near-Infrared Camera), all of the JWST instruments can probe satellites 
well outside Hydra's orbit. Although each instrument has tools to mask
out part of the field, MIRI is not sensitive enough to detect Nix or Hydra
at 10~\mum\ (see below). NIRCam and NIRISS have complementary masking options
and similar sensitivity for K-band imaging, with 10$\sigma$ detections of 
10 nJy sources expected in exposure times of $10^4$~s.

Among the known satellites, Nix and Hydra have V = 23--24 and equivalent
spherical radii of roughly 20~km. Adopting optical/infrared colors of the 
Sun for these high albedo objects \citep[V--K = 1.5;][]{weaver2016}, 
K $\approx$ 22. For point sources having a flux of roughly $10^3$~nJy, a 
100~s integration with a JWST imager yields a 10$\sigma$ detection.  With 
V = 26--27 and K $\approx$ 25, the smaller Styx and Kerberos require a 
factor of 10--20 longer integration time for a 10$\sigma$ detection.

If Nix and Hydra have the mid-IR colors of the Sun, we expect 5~\mum\ (10~\mum) 
fluxes of 200~nJy (60~nJy). At their long-wavelength cutoffs of roughly 4.5~\mum,
NIRCam and NIRISS can achieve S/N = 10 detections in 200~s integrations.  
Using MIRI, we expect S/N = 10 (3) detections of Nix or Hydra with $10^4$~s 
integration times at 5~\mum\ (10~\mum). MIRI's coronagraph is unavailable at 
5~\mum; high quality detections at 10~\mum\ require much longer integration 
times than shorter wavelength observations with either NIRCam or NIRISS.

If there are any 1--2~km high albedo satellites in the \pc\ system, they are 
100--200~times fainter than Nix/Hydra.  Integrations which yield 10$\sigma$ 
detections of Styx and Kerberos result in 3$\sigma$ detections for these 
putative satellites. Longer integrations improve the likelihood of identifying 
any 1--2~km satellites in the system.

Overall, detecting the known satellites of \pc\ with JWST is straightforward. 
A program that acquires 5--10 $10^3$ sec integrations yields excellent S/N for 
Nix/Hydra, very good S/N for Kerberos/Styx, and sufficient S/N to detect 
1--2~km objects from the orbit of Styx to well outside the orbit of Hydra.

\subsection{Occultations}
\label{sec: obs-occ}

The \pc\ system has a long and distinguished history of stellar occultation 
observations \citep[e.g.,][and references therein]{person2013,boissel2014,
gulbis2015,bosh2015,pasachoff2016,pasachoff2017}. Aside from demonstrating 
the presence of Pluto's atmosphere \citep{elliot1989} and measuring physical 
conditions within the atmosphere \citep[e.g.,][]{hubbard1988,elliot2003,
pasachoff2005}, data from occultations provide detailed astrometry on the 
\pc\ orbit, the ephemerides of the \pc\ system, and information on dust and 
other small objects orbiting \pc. Although Charon has been probed with 
occultations \citep[e.g.,][]{gulbis2015}, the lone published attempt to 
monitor an occultation by Nix did not yield the expected drop in stellar 
flux \citep{pasachoff2016}.

For the \pc\ system with a semimajor axis $a \approx$ 40~AU, occultations of 
small satellites with $r \approx$ 1--2~km are in the Fresnel limit 
\citep[e.g.,][and references therein]{bailey1976,cooray2003a,roques2006,
nihei2007,bickerton2008,wang2009,schlicht2009,schlicht2012}. On this scale, 
the duration of an occultation is
\begin{equation}
\Delta t = 2 D \phi_{\star} (1 + A^{1/2}) / v
\end{equation}
where D is the distance from the Earth to Pluto, 
$\phi_\star$ is the angular size of the occulted star,
$v$ is the relative velocity, and
$A = (r/R_\star)^2$ is the amplitude of the occultation.
In the amplitude, $r$ is the radius of the satellite and
$R_\star = D \phi_\star$ is the projected size of the star at Pluto.

Achieving a large amplitude requires that the projected size of the star 
is comparable to the radius of the satellite.  With $R_\star$ = 1--10~km
and $D$ = 40 AU, $\phi_\star \approx 2 - 20 \times 10^{-10}$. If the
physical radius of the star is similar to the solar radius, the distance 
to the star is $d \approx$ 10--100~pc. 

Although the amplitude of the occultation can be large, the duration is
very short. For a 1~km (10~km) satellite with $R_\star$ = 1 km, Pluto's 
orbital motion of $v$ = 5~\kms\ implies $\Delta t$ = 0.8~s (8~s). Resolving 
the event requires integration times of 0.04~s (0.4~s). The relatively long 
duration of occultations of 10--20~km satellites makes the event much easier 
to resolve than occultations by 1--2~km satellites.

Relaxing our assumption of spherical satellites makes occultation observations
somewhat more challenging. From \nh\ data, the aspect ratios range from 
roughly 1.5:1:1 for Nix to roughly 2:1:1 for Styx and Kerberos to roughly
2.6:1.8:1 for Hydra. For a worst-case where long axis of the satellite lies
perpendicular to its path across the star, it is prudent to reduce the 
integration times by a factor of two to $\lesssim$ 0.02~s for Styx/Kerberos 
and by a factor of 2--3 to $\lesssim$0.2~s for Nix/Hydra.

\subsection{Future Options}
\label{sec: obs-fut}

Although ground-based occultations and JWST IR imaging are the only current 
options for detecting smaller satellites in the \pc\ system, we briefly 
consider whether planned missions are also capable of discovering new 
satellites.

WFIRST is a 2.4-m space telescope concept with a wide-field imager and a 
coronagraphic instrument \citep{green2012,spergel2013}. 
Despite having a sensitivity comparable with HST, the wide-field imager is 
not designed for the high contrast, high resolution observations required 
to detect existing or additional satellites in the \pc\ system. The 
coronagraphic instrument is expected to achieve contrast ratios of $10^{-9}$ 
or larger on scales of roughly 1\arcsec\ \citep[see also][]{burrows2014,lacy2018}.
Detecting very small satellites ($r \ll$ 1~km) and any faint debris in the 
system would be straightforward. The likely field-of-view has a radius of 
only 2\arcsec; thus, it would not be possible to center the coronagraph on 
Pluto and search for faint satellites beyond the orbit of Hydra. Still, it
is worth exploring whether moving \pc\ off the edge of the field would allow
deep searches for faint satellites.

The Origins Space Telescope is a 6--9-m telescope concept with 4--5 proposed 
instruments \citep[e.g.,][]{meixner2017,battersby2018}. Designed to operate
at wavelengths $\lambda \gtrsim$ 5~\mum, current plans do not include an 
imaging instrument similar to those on JWST or WFIRST. Unless designs change, 
this facility is unlikely to enable discovery of faint satellites in the 
\pc\ system.

LUVOIR (ATLAST) is a 8--16-m ultraviolet--optical--infrared telescope concept 
with a variety of proposed instruments \citep[e.g.,][]{postman2009,feinberg2014,
thronson2016,bolcar2017,arney2017}. Scaling from quoted sensitivity estimates 
for instruments on the 6.5-m JWST, IR imagers on LUVOIR could detect small 
satellites beyond the orbit of Hydra with radii $r \approx$ 0.5--1~km. 
Improvements in instrument efficiency would enable detection of smaller 
satellites. If the proposed coronagraph could reach a contrast of $10^{-9}$ 
over a larger field-of-view, detection of m-sized objects might be possible.

The Habitable Exoplanet Observatory (HabEx) is a 4--8-m telescope concept 
designed to image exoplanets and to detect biosignatures in the spectra of
exoplanets around nearby stars \citep[e.g.,][]{mennesson2016}. With an 
aperture comparable to JWST and wavelength coverage similar to LUVOIR, 
discovering satellites with $r \lesssim 1$~km requires a coronagraph with 
new technology \citep[e.g.,][]{ruane2017,ruane2018}. As with LUVOIR, 
reaching proposed contrasts of $\sim 10^{-9}$ enables detection of much 
smaller satellites.

\section{DISCUSSION}
\label{sec: disc}

\subsection{Dynamical Architecture of the Pluto--Charon Satellite System}
\label{sec: disc-arch}

The \nbody\ calculations discussed here place new constraints on the 
dynamics of the \pc\ satellite system. For the adopted masses of the four 
small satellites, the system is as tightly packed as possible. There are
few stable orbits where satellites with negligible mass can exist between the 
orbit of Styx and the orbit of Hydra. This conclusion is independent of 
inclination angle: satellites on polar orbits are about as unstable as prograde 
satellites orbiting in the plane of the binary system. 

There is also limited space for stable satellites on circumbinary orbits inside 
the orbit of Styx\footnote{For discussions of stability regions at smaller $a$
(between Pluto and Charon), see \citet{winter2010} and \citet{giuliatti2013,
giuliatti2014,giuliatti2015}.} \citep[see also][]{holman1999,pichardo2008,
doolin2011,youdin2012}.  From simulations without any small satellites, stable 
orbits could exist from 1.3--2.1~\apc\ (retrograde) and 1.7--2.1~\apc\ (prograde). 
Polar orbits inside the orbit of Styx are unstable.  The new calculations yield 
a set of massless survivors on prograde orbits with initial semimajor axes 
$a_0 \approx$ 1.8--2.1~\apc.

Well outside the orbit of Hydra, massless satellites on prograde and polar orbits 
are stable.  With the nominal masses, Nix and Hydra clear out tracers with 
$a \lesssim 1.08~a_H$ on time scales of 5--10~Myr. On longer time scales, orbits 
with $a \approx 1.08-1.12~a_H$ are probably also unstable. Tracers with 
$a \gtrsim 1.15~a_H$ appear to be stable. 

Tests with massive satellites confirm these results. Although some massless tracers 
on prograde orbits survive for as long as 10~Myr near the corotation zones of Nix 
and Hydra, small satellites with $r \approx$ 2~km and $m \approx 4 \times 10^{16}$~g 
on identical orbits have lifetimes as short as 10--1000~yr. Others survive for 100~Myr.
Inside the orbit of Styx, small moons are ejected on time scales ranging from 1~yr 
to nearly 100~Myr.  Small moons on polar orbits between Nix and Hydra are similarly
unstable, with lifetimes ranging from $10^3$~yr to 100~Myr.  Given the loss rate in
our calculations, it seems unlikely that additional satellites with $r \gtrsim$ 2~km
inside the orbit of Hydra can survive for the age of the solar system.

Massive satellites outside the orbit of Hydra fare much better. After 150~Myr of
evolution, roughly 90\% of those on prograde orbits survive. Survivors tend to have 
larger initial semimajor axes than ejected moons. Small moons on polar orbits outside
the orbit of Hydra are just as unlikely to be ejected after 100~Myr of evolution.
Thus, searches for new satellites should concentrate on regions outside the orbit
of Hydra.

Aside from clarifying our understanding of circumbinary dynamics, these results
help to interpret the lack of new satellite detections from \nh\ imaging data. 
With additional satellites generally precluded at $a \lesssim 1.1~a_H$, the 
volume for discovering new, stable satellites within the overall \nh\ footprint 
$a \lesssim 1.6~a_H$ is restricted. While it is not surprising that \nh\ failed
to detect new satellites inside the orbit of Hydra, the lack of 2--3~km satellites
outside the orbit of Hydra is surprising. Perhaps they never formed or were ejected.
If smaller than the \nh\ threshold of 2~km, they await detection with a new generation
of instruments.

The calculations of massless tracers orbiting between Styx and Hydra also provide 
new insights into the lack of emission from small particles detected from \nh\ data 
\citep{lauer2018}. Prior to the \nh\ flyby, several studies derived upper limits
on small moons and dust emission from direct imaging \citep[e.g.,][]{steffl2006,
marton2015} and occultations \citep{boissel2014,throop2015}. Theoretical studies 
based on \nbody\ simulations predicted steady-state optical depths from a balance 
between dust production from impacts on \pc\ and the smaller satellites and losses 
from radiation pressure and dynamical ejections \citep{stern2006,pires2011,
poppe2011,pires2013}. The new results described here demonstrate that all orbits 
from 1.7~\apc\ to 3.6--3.7~\apc\ are dynamically unstable on time scales ranging 
from several decades to 10~Myr. Without additional dust production from impacts, 
our results imply that there should be no dust or larger particles inside the orbit 
of Hydra.  

If small particles or dust emission are ever detected in the \pc\ system, the mass 
and location of small particles place interesting constraints on the masses of the 
four small satellites. If limits on the production rate for small particles from 
impacts can be established from \nh\ data, the number of survivors between the
orbits of Styx and Hydra is sensitive to the mass of Nix and Hydra: smaller masses
for these satellites allow larger masses in dust \citep[see also][]{stern2006,
pires2011,poppe2011,pires2013}. 

Identifying new satellites or small particles outside the orbit of Hydra would also 
improve estimates for the mass of Hydra. From the \nbody\ calculations, the innermost
stable orbit outside Hydra is much more sensitive to Hydra's mass than to the mass of
Nix or the other satellites \citep[see also][]{michaely2017}. Accurate measurements of
the orbital elements for any new satellite would thus provide new limits on Hydra's mass.

\subsection{JWST Feasibility}
\label{sec: disc-jwst}

Although the current launch date for JWST is not until 2021 March, the observations
proposed here are similar to the suite of HST imaging data collected prior to the \nh\ flyby
\citep[e.g.,][]{weaver2006,buie2006,steffl2006,steffl2007,tholen2008,showalter2011,
showalter2012,brozovic2015,showalter2015}. Starting in Cycle 2, various HST imaging 
programs sought to constrain the properties of dusty structures and satellites in the 
\pc\ system. These programs acquired many images per HST orbit, with exposure times
ranging from a few seconds to several minutes. As outlined in \citet{brozovic2015}, 
multiple images per orbit enable a robust analysis procedure which eliminates light
from background stars and provides the best possible signal-to-noise for images of 
very faint satellites.

We anticipate that a JWST observing program to detect faint satellites in the \pc\ system
would be similar to a typical Hubble program. Within a single visit, the likely total 
exposure time with JWST imagers would probably be several times longer than an HST orbit.
As outlined in \citet{gordon2015} for MIRI, multiple short and moderate exposures would 
allow the same type of analysis procedure as performed by \citet{brozovic2015}.

Aside from the ability of JWST instruments to perform to specifications, the main 
uncertainty in any JWST program is the overhead involved in acquiring a target, 
maintaining a fix on the target, conducting the observations, and performing the
housekeeping needed for the health of the satellite. Various performance analyses
suggest the typical overhead in an observing program is roughly 30\% \citep{gordon2012a,
gordon2012b}. Thus, the program outlined here seems feasible.

\subsection{Occultation Feasibility}
\label{sec: disc-occs}

Over the past few decades, various groups have observed Pluto and Charon 
occult fairly bright stars to infer the extent and physical properties of 
their atmospheres and to plan for the \nh\ flyby \citep[e.g.,][and references 
therein]{person2006,sicardy2011,person2013,throop2015,gulbis2015,bosh2015,
dias2015,sicardy2016,pasachoff2016,pasachoff2017}. 
With exposure times of 0.25 to 5 sec on 1-m to 2.5-m telescopes, the 
typical signal-to-noise ranges from roughly 10 to better than 100. 

Detailed astrometric analyses demonstrate that occultations by the \pc\ system
are fairly frequent. For 2008--2015, predictions by \citet{assafin2010} indicate
30--300 possible occultations per year for each of Pluto, Charon, Nix, and Hydra.
Although we are unaware of similar published predictions for Styx and Kerberos, it 
seems likely from \citet{assafin2010} that the frequency of occultations for both 
of these small moons is similar to that for Nix and Hydra.  The uncertain orbital 
path of Pluto across the sky is the main limitation in these predictions
\citep[e.g.,][and references therein]{benedetti2014,holman2016}. 

Many studies have explored the possibility of using occultations to detect 
small Kuiper belt objects beyond 40--50 AU \citep[KBOs; e.g.,][and references 
therein]{bailey1976,brown1997,alcock2003,cooray2003a,cooray2003b,chang2006,
roques2006,bickerton2008,bianco2009,schlicht2009,wang2009,bianco2010,wang2010,
chang2011,schlicht2012,zhang2013}.  Aside from ground-based optical observations, 
HST and RXTE have yielded promising sets of data to search for serendipitous 
occultations of stellar sources by KBOs.  On the ground and with HST, exposure 
times of 0.02 sec yield light curves with sufficient signal-to-noise to detect 
KBOs with radii of roughly 1~km.

Based on this discussion, it seems straightforward to detect occultations of 
stars by Pluto's known small satellites with modest aperture ground-based 
telescopes. However, the likelihood of detecting a serendipitous occultation 
event from an unknown satellite is small. Assuming a satellite with a diameter 
$D_6 \approx$ 2~km is positioned randomly in an orbit with semimajor axis 
$a_6 \approx$ 75,000~km (outside Hydra), the chance of having it along the 
same occultation path as any of the other small satellites is less than 
$2 D_6 / 2 \pi a_6$ $\lesssim 10^{-5}$. With $N$ of these satellites, the 
probability is still small, $10^{-5} N$, unless $N$ is very large.

Overall, the best strategy to detect and to characterize new satellites involves 
initial imaging observations with JWST followed by ground-based occultation 
observations. Although JWST data are fairly expensive and limited by spacecraft 
constraints, observations with any of the imaging instruments sample broad swaths 
of available discovery space.  If new satellites are detected, occultations can
then provide additional information on the orbits and shapes.

\subsection{Connections with Exo-Planetary Systems}

Over the past few decades, various techniques have revealed $\sim$ 15 circumbinary
planetary systems \citep{thorsett1993,doyle2011,welsh2012,orosz2012a,orosz2012b,
schwamb2013,kostov2013,bailey2014,kostov2014,kostov2016,jain2017,getley2017}. 
Aside from systems with main sequence stars, the binaries include a binary pulsar
and a low mass X-ray binary. Planet masses range from $\sim$ Neptune up to $\sim$ 
seven times Jupiter.  Often, the circumbinary planets orbit in the plane of the 
inner binary and are reasonably close to the innermost stable orbit. In several, 
the orbit of the planet is somewhat tilted with respect to the inner binary. 
Sometimes, the planet is well outside the innermost orbit. 

The formation and evolution of circumbinary planets have attracted intense 
theoretical interest \citep[e.g.,][and references therein]{pierens2007,pierens2008a,
pierens2008b,paarde2012,raf2013,pierens2013,rafikov2015a,rafikov2015b,silsbee2015a,
silsbee2015b,bk2015a,kley2015,vartanian2016,hamers2016,li2016,mutter2017a,mutter2017b,
quarles2018b,fleming2018,pierens2018,zanazzi2018,thun2018}. In addition to changing 
the structure and evolution of a planet-forming gaseous disk, the central binary 
provides a challenging environment for the growth of Earth-mass and larger planets 
from km-sized and larger planetesimals. Once planets form, the central binary 
efficiently removes them from resonant orbits. Multi-planet systems are particularly
prone to disruption.

Although there are currently no circumbinary planetary systems with more than one 
planet, there are numerous multi-planet systems orbiting single stars 
\citep[e.g.,][and references therein]{lissauer2011b,lissauer2012,rowe2014,fabrycky2014,
lissauer2014,winn2015,ballard2016,sinukoff2016,udry2017,weiss2018}.
Many of these are closely-packed, with little or no space for additional planets 
between the innermost and outermost planets. In each multi-planet system, the planets
are often of similar size, with the outermost planet usually the largest \citep[see 
also][]{ciardi2013,milholland2017}.  Sometimes, the planets have very different sizes. 

Theoretical studies of closely-packed multi-planet systems focus on their origin 
and stability \citep[e.g.][and references therein]{rein2012,hansen2012,hansen2013,
raymond2014,schlaufman2014,najita2014,steffen2015,malhotra2015,pu2015,batygin2015a,
morrison2016,pan2017,mustill2017}. Formation at several AU distances followed by 
migration through a circumstellar gaseous disk is a favored explanation for many 
close-in multi-planet systems. However, {\it in situ} growth is also a possibility.
In both mechanisms, it is unclear how the most closely-packed planetary systems 
form and maintain their stability on Gyr time scales.

Together with previous theoretical studies of the \pc\ system \citep{canup2005,
ward2006,lee2006,lith2008a,lith2008b,canup2011,youdin2012,kb2014b,cheng2014b,
desch2015,pires2015, walsh2015,bk2015b,michaely2017,smullen2017,mckinnon2017,
woo2018}, the results described here inform our understanding of circumbinary 
planets and closely-packed planetary systems orbiting single or binary stars. 
Despite forming in a relatively gas-free environment, the \pc\ system has issues 
with {\it in situ} formation, migration, orbital resonances, and long-term 
stability similar to those in exoplanetary systems. Nevertheless, the small 
satellites probably (i) grew from a ring of debris, (ii) found stable orbits
close to resonances with the central binary, and (iii) maintained these orbits
for $\sim$ 4~Gyr \citep{weaver2016,robbins2017}.  Working out the details of 
this history for the \pc\ satellites and exoplanetary systems will enrich 
theories of planet formation. 

\section{SUMMARY}

We have analyzed a large suite of \nbody\ calculations to isolate stable orbits
for additional satellites in the \pc\ system. Although there are few stable low 
inclination, prograde orbits or high inclination, polar orbits for massless tracers
with semimajor axes, $0.9 ~ a_S \lesssim a \lesssim 1.1 ~ a_H$, low eccentricity
orbits with $a \gtrsim 1.1 ~ a_H$ are stable. Within this range of semimajor axes,
polar and prograde orbits are equally stable. 

Tests with massive satellites ($r \approx$ 2~km) confirm the stability of orbits 
well beyond the orbit of Hydra. Among an ensemble of satellites with $a_0 \approx$ 
3.7--4.0~\apc, nearly all survive 100--150~Myr of dynamical evolution. Calculations 
of satellites with smaller $a_0$ have many fewer survivors after 100--150~Myr. Thus,
the best region to search for new satellites in the \pc\ system is beyond the orbit 
of Hydra.

Several types of observations could detect small satellites on these orbits. Direct
imaging with JWST can reveal 1--2~km satellites with modest integration times. 
Although discovering such small satellites with stellar occultations is a challenge, 
observations with 2--3-m class telescopes can detect the signal from the occultation 
of a nearby solar-type star by a 1--2~km satellite. In the (far) future, observations 
with WFIRST or ATLAST may reveal even smaller satellites and a debris disk or ring(s).

Finding additional small satellites in the \pc\ system constrains the masses of the
known satellites and provides additional tests of theoretical models for satellite
formation. Discovery of small objects between the orbits of Styx and Hydra would 
require lower mass (and mass density) for Nix and Hydra. Any satellite orbiting 
beyond Hydra might reduce the uncertainty in the mass of Hydra. Current theory 
predicts several satellites with $r \approx$ 1--3~km and $a \approx 1.5-2.5 ~ a_H$. 
Observations with JWST and ground-based telescopes can test this theory and improve 
our understanding of circumbinary satellite formation.

\vskip 6ex

Resources supporting this work on the `discover' cluster were provided by the NASA 
High-End Computing (HEC) Program through the NASA Center for Climate Simulation 
(NCCS) at Goddard Space Flight Center.  Advice and comments from T. Currie, 
M. Geller, M. Payne, and A. Youdin greatly improved our presentation.  
Portions of this project were supported by the {\it NASA } {\it Outer Planets} 
and {\it Emerging Worlds} programs through grants NNX11AM37G and NNX17AE24G.

\vskip 6ex

\appendix

\section{Tests of the Symplectic Integrator}

To track the orbital evolution of the \pc\ system, our \nbody\ code employs an adaptive 
sixth-order accurate algorithm based on either Richardson extrapolation \citep{bk2006} or 
a symplectic method \citep{yoshida1990,wisdom1991,saha1992}.  The code calculates 
gravitational forces by direct summation and evolves particles in the center-of-mass 
frame. Aside from passing a stringent set of dynamical tests and benchmarks 
\citep{dunc1998,bk2006}, we have used the code to simulate scattering of super-Earths by 
growing gas giants \citep{bk2011a}, migration through planetesimal disks \citep{bk2011b} 
and Saturn's rings \citep{bk2013}, the formation of Pluto's small satellites \citep{kb2014a}, 
and the circularization of the orbits of planet scattered into the outer solar system 
\citep{bk2014,bk2016}.

To evolve the orbit of the \pc\ satellites in time, the time step in our symplectic 
algorithm is $\Delta t = T_{PC} / N$ where $T_{PC}$ is the orbital period of the 
central binary and $N$ is an integer. For any simulation, the total cpu time is 
proportional to $N$. To select a value for $N$ which maintains the integrity of the
solution in a reasonable amount of cpu time, we consider the orbit of an idealized 
\pc\ binary with the measured masses and orbital semimajor axes and orbital 
eccentricity $10^{-4}$, $10^{-5}$, $10^{-6}$, and $10^{-7}$.  For $N$ = 20--150, we 
evolve the binary orbit for 100~Myr and record the position $(x, y, z)$ and velocity 
$(\dot{x}, \dot{y}, \dot{z})$ vectors and the osculating orbital elements $a$ and $e$
every 10--100 binary orbits.  To evaluate the ability of the code to track $a$ and $e$, 
we derive the average, standard deviation, median, and inter-quartile range over $M$ 
time steps. Typically, the median is indistinguishable from the average; the 
inter-quartile range is nearly identical to the standard deviation. Using standard 
estimates for the linear correlation coefficient (Pearson's $r$), the Spearman rank-order 
correlation coefficient, and Kendall's $\tau$ \citep{press1992}, we look for trends in 
$a$, $e$, and $\imath$ with evolution time.

In these tests, there is no indication that the average/median $a$ (and its standard
deviation or inter-quartile range) or any trend in $a$ and $e$  with time depend on the
number of steps per binary orbit. For the semimajor axis, the vanishingly small dispersion
and inter-quartile range are independent of $M$. Typical correlation coefficients of $a$
and $e$ with time are $\pm 10^{-3}$ or smaller. Thus, none of our calculations experience
any drift in $a$ or $e$ over 100~Myr of evolution.

\begin{figure}
\includegraphics[width=6.5in]{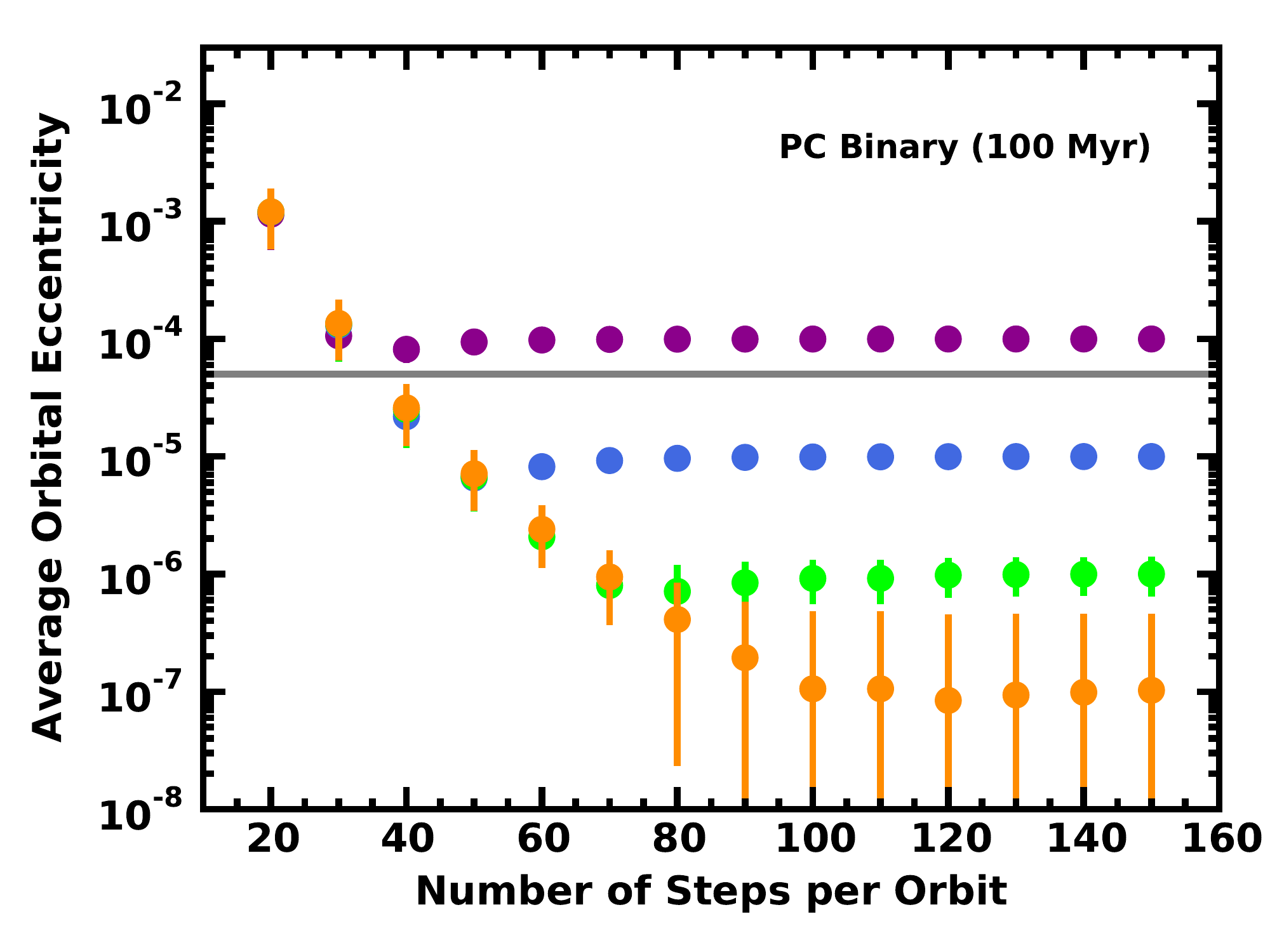}
\vskip 2ex
\caption{\label{fig: e-test}
Average eccentricity (symbols) and standard deviation (bars) of the \pc\ binary over 
100~Myr of evolution as a function of $N$ the number of symplectic steps per binary 
orbit. Colors indicate the input eccentricity: $e_0 = 10^{-4}$ (purple), 
$e_0 = 10^{-5}$ (blue), $e_0 = 10^{-6}$ (green), and $e_0 = 10^{-7}$ (orange).
Calculations with $N \ge$ 40 maintain $e$ near or below the measured
$e = 5 \times 10^{-5}$ indicated by the horizontal grey line. Smaller $N$ generates 
orbital $e$ much larger than the observed $e$. The larger standard deviation for 
$e \approx 10^{-7}$ is due to round-off error in our method for deriving $e$ from 
the orbital position and velocity.
}
\end{figure}

Trends of the average $e$ and the standard deviation in $e$ with $N$ are very clear
(\ref{fig: e-test}). For the $e = 10^{-7}$ binary (orange symbols), calculations with 
$N \ge$ 100 reliably maintain the initial orbital configuration. Results for $N$ = 
20--30 are especially poor. When $e = 10^{-6}$ (green symbols), calculations with 
$N \ge$ 70 sustain the initial $e$. Faithfully tracking the orbits of binaries with 
larger $e$ requires fewer steps per binary orbit, $N$ = 30 for $e = 10^{-4}$ (purple 
symbols) and $N \approx$ 40--50 for $e = 10^{-5}$ (blue symbols). With a measured
$e \approx 5 \times 10^{-5}$ (as indicated by the horizontal grey line), calculations
with $N$ = 40 preserve the measured $e$ of \pc.

Long (100--500~Myr) simulations of \pc\ and the four satellites yield similar results 
for trends in $a(t)$ and $e(t)$ of the central binary. As long as the satellite system
remains stable, there is no trend in $a$ or $e$ of the Pluto-Charon binary or the small
satellites. We plan to describe the results of these simulations in a separate publication. 

%\bibliography{sfpl}
\bibliography{ms.bbl}

\clearpage

\begin{deluxetable}{lccccccccc}
\tablecolumns{9}
\tablewidth{0pc}
\tabletypesize{\scriptsize}
\tablenum{1}
\tablecaption{Nominal satellite properties for \nbody\ calculations\tablenotemark{a}}
\tablehead{
  \colhead{Satellite} &
  \colhead{$m_s$} &
  \colhead{$r_s$ (km)} &
  \colhead{$\rho_s$} &
  \colhead{$\rhill$ (km)} &
  \colhead{$a$ (km)} &
  \colhead{$a / a_{PC}$} &
  \colhead{$e$ ($\times 10^{-3}$)} &
  \colhead{$\imath$ (deg)} &
  \colhead{$P_{orb}$ (d)}
}
\startdata
Styx & 0.60 & 5.2 & 1.00 & 198 & 42,656 & 2.263 & 5.787 & 0.809 & 20.16155 \\
Nix & 45.0 & 20.0 & 1.34 & 487 & 48,694 & 2.583 & 2.036 & 0.133 & 24.85463 \\
Kerberos & 1.0 & 6.0 & 1.11 & 405 & 57,783 & 3.065 & 3.280 & 0.389 & 32.16756 \\
Hydra & 48.0 & 20.0 & 1.24 & 661 & 64,738 & 3.434 & 5.862 & 0.242 & 38.20177\\
\enddata
\label{tab: sats}
\tablenotetext{a}{Based on published analyses of the 
mass ($m_s$ in units of $10^{18}$ g), radius ($r_s$), 
density ($\rho_s$ in units of g~cm$^{-3}$), Hill radius (\rhill),
semimajor axis ($a$), orbital eccentricity ($e$) and inclination ($\imath$),
and orbital period \citep[$P_{orb}$][]{brozovic2015,stern2015,weaver2016,
nimmo2017,mckinnon2017}.
}
\end{deluxetable}
\clearpage

\begin{deluxetable}{lccccc}
\tablecolumns{8}
\tablewidth{5in}
\tabletypesize{\scriptsize}
\tablenum{2}
\tablecaption{Survivor fraction for \nbody\ calculations\tablenotemark{a}}
\tablehead{
  \colhead{Orbit} &
  \colhead{Satellites} &
  \colhead{Tracer Mass} &
  \colhead{~~~~~~$a_0 / a_{PC}$~~~~~~} &
  \colhead{~~~~~~~~~~$f_s$}~~~~~~~~~~ &
  \colhead{Time Scale}
}
\startdata
Prograde & No & No & 1.00--1.50 & 0.00 & 10 Myr \\
Prograde & No & No & 1.45--2.10 & 0.29 & 10 Myr \\
Prograde & No & No & 1.95--2.65 & 0.90 & 10 Myr \\
Prograde & No & No & 2.60--3.25 & 1.00 & 10 Myr \\
Retrograde & No & No & 1.00--1.50 & 0.20 & 10 Myr \\
Retrograde & No & No & 1.45--2.10 & 0.98 & 10 Myr \\
Retrograde & No & No & 1.95--2.65 & 1.00 & 10 Myr \\
Retrograde & No & No & 2.60--3.25 & 1.00 & 10 Myr \\
Polar & No & No & 1.00--1.50 & 0.00 & 10 Myr \\
Polar & No & No & 1.45--2.10 & 0.00 & 10 Myr \\
Polar & No & No & 1.95--2.65 & 0.02 & 10 Myr \\
Polar & No & No & 2.60--3.25 & 0.10 & 10 Myr \\
\\
Prograde & Yes & No & 1.60--2.10 & 0.19 & 10 Myr \\
Prograde & Yes & No & 2.10--2.60 & 0.02 & 10 Myr \\
Prograde & Yes & No & 2.40--3.00 & 0.05 & 10 Myr \\
Prograde & Yes & No & 2.80--3.40 & 0.05 & 10 Myr \\
Prograde & Yes & No & 3.20--4.00 & 0.43 & 10 Myr \\
Polar & Yes & No & 1.60--2.10 & 0.00 & 10 Myr \\
Polar & Yes & No & 2.10--2.60 & 0.00 & 10 Myr \\
Polar & Yes & No & 2.40--3.00 & 0.21 & 10 Myr\tablenotemark{b} \\
Polar & Yes & No & 2.80--3.40 & 0.14 & 10 Myr\tablenotemark{c} \\
Polar & Yes & No & 3.20--4.00 & 0.60 & 10 Myr\tablenotemark{d} \\
\\
Prograde & Yes & Yes & 1.76--2.02 & 0.10 & 100 Myr\tablenotemark{e}\\
Prograde & Yes & Yes & 2.48--2.54 & 0.10 & 100 Myr\tablenotemark{f}\\
Prograde & Yes & Yes & 3.29--3.36 & 0.21 & 100 Myr\tablenotemark{g}\\
Prograde & Yes & Yes & 3.84--3.97 & 0.96 & 100 Myr\tablenotemark{h}\\
Polar & Yes & Yes & 2.67--2.80 & 0.39 & 100 Myr \\
Polar & Yes & Yes & 3.00--3.09 & 0.00 & 100 Myr \\
Polar & Yes & Yes & 3.71--3.91 & 1.00 & 100 Myr \\
\enddata
\label{tab: fracs}
\tablenotetext{a}{The first column lists the initial sense of the 
orbits for systems of massless tracers (`No' in column three) or one
low mass satellite (`Yes' in column three) with the range of semimajor 
axes in column four. Calculations have the \pc\ binary as the 
central mass, with (`Yes' in column two) or without (`No' in 
column two) the four small satellites. The survivor fraction $f_s$ 
(length of the simulation) is listed in column five (six).}
\tablenotetext{b}{After 20~Myr, the survivor fraction is 0.18}
\tablenotetext{c}{After 20~Myr, the survivor fraction is 0.08}
\tablenotetext{d}{After 20~Myr, the survivor fraction is 0.57}
\tablenotetext{e}{After 150~Myr, the survivor fraction is 0.03}
\tablenotetext{f}{After 150~Myr, the survivor fraction is 0.10}
\tablenotetext{g}{After 150~Myr, the survivor fraction is 0.14}
\tablenotetext{h}{After 150~Myr, the survivor fraction is 0.90}
\end{deluxetable}
\clearpage

\clearpage

\clearpage

\clearpage

\clearpage

\clearpage

\clearpage

\clearpage

\clearpage

\clearpage

\clearpage

\clearpage

\clearpage

\end{document}